\documentclass[12pt]{article} 

\usepackage{amsmath}
\usepackage{natbib}
\usepackage{graphicx}
\usepackage{psfrag}
\usepackage{a4wide}
\usepackage{algorithmic}
\usepackage{algorithm}

\title{Patterns in high-frequency FX data: \\ Discovery of 12 empirical scaling laws}

\author{
{J.B.~Glattfelder$^{\ast \dag}$, A.~Dupuis$^{\dag \ddagger}$ and R.B.~Olsen$^{\dag \ddagger}$}\\
{\footnotesize ${\dag}$ Olsen Ltd., Seefeldstrasse 233, 8008 Zurich, Switzerland}\\
{\footnotesize  $\ddagger$ Centre for Computational Finance and Economic Agents (CCFEA), University of Essex, UK}
}

 \date{}

\begin{document}


\maketitle

\begin{abstract}
We have discovered 12 independent new empirical scaling laws in
foreign exchange data-series that hold for close to three orders of
magnitude and across 13 currency exchange rates. Our statistical
analysis crucially depends on an event-based approach that measures
the relationship between different types of events. The scaling laws
give an accurate estimation of the length of the price-curve
coastline, which turns out to be surprisingly long. The new laws
substantially extend the catalogue of stylised facts and sharply
constrain the space of possible theoretical explanations of the market
mechanisms.
\end{abstract}

\renewcommand{\thefootnote} {\fnsymbol{footnote}}
\footnotetext[1]{Corresponding author. Email: jbg@olsen.ch.}

\section{Introduction}

The global financial system has recently been rocked by losses that
could total four trillion USD (\cite{imf:09}). The crisis is seriously
undermining the functioning of the financial system, the backbone of
the global economy.  This suggests an acute deficiency in our
understanding of how markets work. Are there ``laws of nature'' to be
discovered in financial systems, giving us new insights? We approach
this question by identifying key empirical patterns, namely
scaling-law relations. We believe that these universal laws have the
potential to significantly enhance our understanding of the markets.

Scaling laws establish invariance of scale and play an important role
in describing complex systems (e.g. \cite{west:97,barabasi:99,newman:05}). 
In finance, there is one scaling law that has been widely reported
(\cite{muller:90, mantegna:95, galluccio:97, guillaume:97, ballocchi:99,
dacorogna:01, corsi:01,dimatteo:05}): the size of the average absolute
price change (return) is scale-invariant to the time interval of its
occurrence. This scaling law has been applied to risk management and
volatility modelling (see \cite{ghashghaie:96,gabaix:97,sornette:00,
dimateo:07}) even though there has been no consensus amongst
researchers for why the scaling law exists (e.g., \cite{bouchaud:00,
barndorff-nielsen:01,farmer:04,lux:06, joulin:08}).

In the challenge of identifying new scaling laws,
we analyse the price data of the foreign exchange (FX) market, a
complex network made of interacting agents: corporations,
institutional and retail traders, and brokers trading through market
makers, who themselves form an intricate web of interdependence.  With
a daily turnover of more than 3 trillion USD (\cite{bis:07})  and with price
changes nearly every second, the FX market offers a unique opportunity
to analyse the functioning of a highly liquid, over-the-counter market
that is not constrained by specific exchange-based rules. In this
study we consider five years of tick-by-tick data for 13 exchange
rates through November 2007 (see section \ref{data} for a description
of the data set).

It is a common occurrence for an exchange rate to move by $10$ to
$20\%$ within a year. However, since the seminal work of
\cite{mandelbrot:63} we know about the fractal nature of
price curves. The coastline, roughly being the sum of all price moves
of a given threshold, at fine levels of resolution may be far longer
than one might intuitively think. But how many times longer? The
scaling laws described in this paper provide a surprisingly accurate
estimate and highlight the importance of not only considering tail
events (\cite{sornette:02}), but set these in perspective with the
remarkably long coastline of price changes preceding them.  It should
be noted that our study is not related to the analysis of lead-lag
relationships.

The remainder of the paper is organised as follows. Our main results are
presented in section \ref{res}. We start by enumerating the empirical 
scaling laws, then cross-check our results by establishing quantitative 
relations amongst them and discuss the coastline. In section \ref{meth} the 
methods and the data are described and we conclude with some final remarks 
in section \ref{conc}.  Finally, appendix \ref{app} contains tables with all 
the estimated scaling-law parameters.

\section{The laws and beyond}
\label{res}

\subsection{The new scaling laws}
\label{newsl}
 
Interest in scaling relations in FX data was sparked in 1990 by
a seminal paper relating the mean absolute change of the 
logarithmic mid-prices, sampled at time intervals $\Delta t$ 
over a sample of size $n \Delta t$, to the size of the time interval 
(\cite{muller:90})
\begin{equation}
\tag{0a}
\label{sldxdt}
\langle |\Delta \chi | \rangle_p = \left( \frac{ \Delta t }
    { C_\chi  (p)} \right)^{E_\chi (p)},
\end{equation}
where $\Delta \chi_i = \chi_{i}-\chi_{{i-1}}$ and $\chi_{i} =
\chi(t_i) = (\ln \mathrm{bid}_i + \ln \mathrm{ask}_i) / 2$ is the
logarithmic mid-price of a currency pair at time $t_i$, and
$E_\chi(p)$, $C_\chi(p)$ are the scaling-law parameters. The averaging
operator is $\langle x \rangle_p = \left( 1/n \sum_{j=1}^n x_{j}^p
\right)^{1/p}$, usually with $p \in \{1,2\}$, and $p$ is omitted if
equal to one. Note that for law (\ref{sldxdt}) the data is sampled at
fixed time intervals $t_i = i \Delta t$. This requires a time
interpolation scheme (described in section \ref{data}) which we will
also employ when necessary in the following. Throughout the paper we
consider a simpler definition of the price given by $x_i=
(\mathrm{bid}_i + \mathrm{ask}_i) /2$ where price moves are defined as
$\Delta x_i = (x_{i}-x_{i-1}) / x_{i-1}$. Although the definition of
$x_i$ looses the mathematical feature of $\chi_i$ of behaving
anti-symmetrically under price inversions (e.g.,
$\chi_i^{\textrm{EUR-USD}} = - \chi_i^{\textrm{USD-EUR}}$) it is more
natural as, practically, percentages are more intuitive to manipulate
than differences between logarithmic values. However, considering
either $\chi_i$ or $x_i$ leads to very similar results even for large
spread values. Note that the logarithmic mid-price is equivalent to
the logarithm of the geometric mean of the bid and ask prices $\chi_i
= \ln \sqrt{\mathrm{bid}_i \ \mathrm{ask}_i}$, whereas $x_i$ is the
arithmetic mean.

Later, in 1997, a second scaling law was reported by
\cite{guillaume:97}, relating the number $\textsf{N}(\Delta
\chi_{dc})$ of so-called directional changes to the the
directional-change sizes $\Delta \chi_{dc}$
\begin{equation}
\tag{0b}
\label{slndx}
\textsf{N} (\Delta \chi_{dc}) = \left( \frac{ \Delta \chi_{dc}}
{ C_{\textsf{\tiny N},dc} } \right)^{E_{\textsf{\tiny N},dc}}.
\end{equation}
%

\begin{figure}[t!]
\centering
\includegraphics[width=0.5\textwidth]{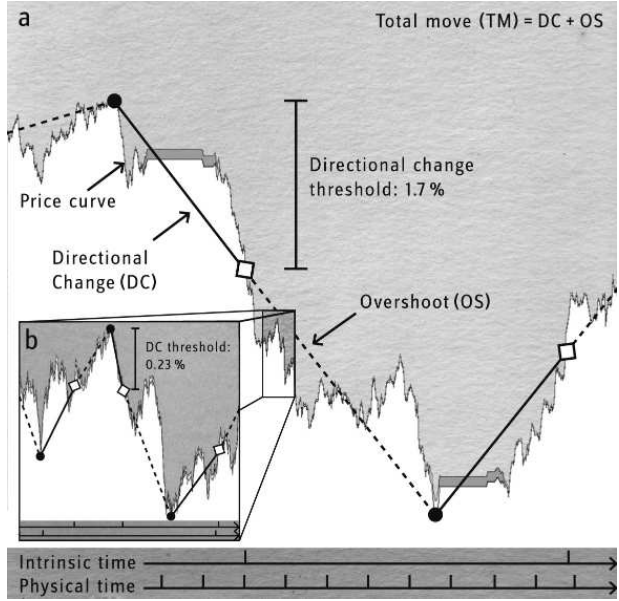}
\caption{
Projection of a (a) two-week, (b) zoomed-in 36 hour price sample onto
a reduced set of so-called directional-change events defined by a
threshold (a) $\Delta x_{dc} = 1.7\%$, (b) $\Delta x_{dc} = 0.23\%$.
These directional-change events (diamonds) act as natural dissection
points, decomposing a total-price move between two extremal price
levels (bullets) into so-called directional-change (solid lines) and
overshoot (dashed lines) sections. The directional-change computation
is detailed in algorithm \ref{alg:dc} of section \ref{pseudo}. Note
the increase of spread size during the two weekends with no price
activity. Time scales depict physical time ticking evenly across
different price-curve activity regimes, whereas \textit{intrinsic
time} triggers only at directional-change events, independent of the
notion of physical time.}
\label{fig:dc_to}
\end{figure}

In financial markets, the flow of time is discontinuous: over weekends
trading comes to a standstill or, inversely, at news announcements
there are spurts of market activity. In law (\ref{sldxdt}), the
confinement of analysing returns as observed in physical time is
overly restrictive. Law (\ref{slndx}) is a first attempt at
establishing a new paradigm by looking beyond such constraints within
financial data, constituting an event-driven approach, where patterns
emerge for successions of events at different magnitudes. This
alternative approach defines an activity-based time-scale called
intrinsic time.

\begin{figure}[h]
\centering
\includegraphics[width=0.8\textwidth]{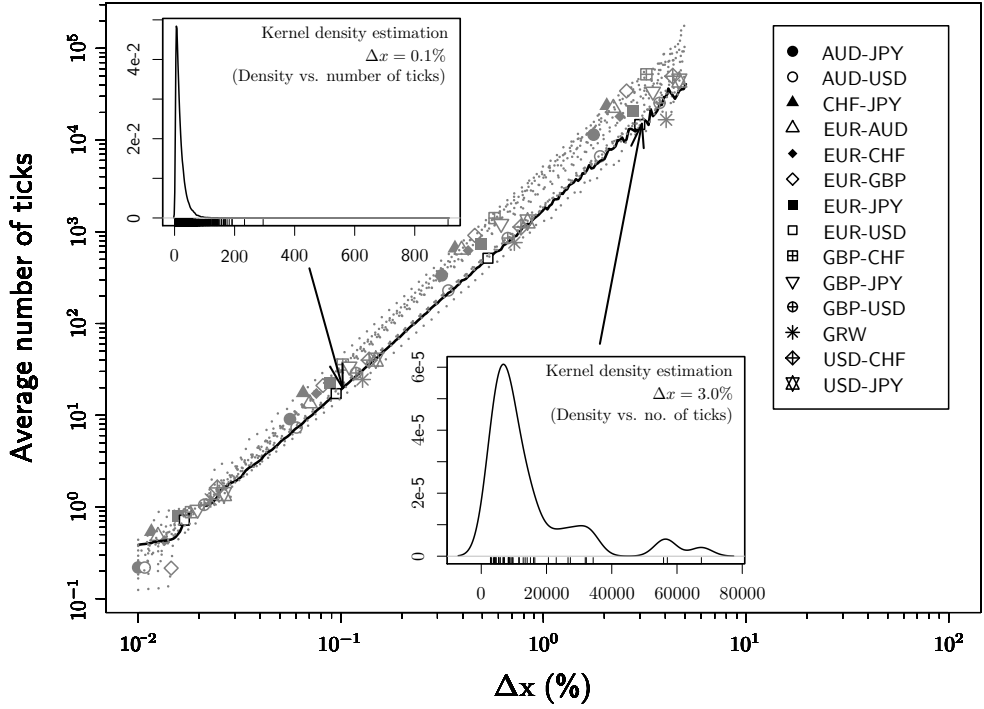}
\caption{Scaling law (\ref{slticks}) is plotted where the x-axis
shows the price move thresholds of the observations and the y-axis
the average tick numbers.  A tick is defined as a price move of $0.02\%$. 
The solid line shows the raw data for EUR-USD. For the remaining 12 
currency pairs and the Gaussian random walk benchmark model the 
raw data is displayed with dots. Insets show the distribution of the 
EUR-USD observations (drawn above their x-axis) for selected threshold 
values of $0.1\%$ and $3.0\%$.  See appendix \ref {app} for the values 
of the estimated scaling-law parameters.
}
\label{fig:slticks}
\end{figure}

Extending this event-driven paradigm further enables us to observe
new, stable patterns of scaling and reduces the level of complexity of
real-world time series. In detail, the fixed event thresholds of
different sizes define focal points, blurring out irrelevant details
of the price evolution. Figure \ref{fig:dc_to} depicts how the price
curve is dissected into so-called directional-change and overshoot
sections. The dissection algorithm measures occurrences of a price
change $\Delta x_{dc}$ from the last high or low (i.e., extrema), if
it is in an up or down mode, respectively. At each occurrence of a
directional change, the overshoot associated with the previous
directional change is determined as the difference between the price
level at which the last directional change occurred and the extrema,
i.e., the high when in up mode or low when in down mode. The high and
low price levels are then reset to the current price and the mode
alternates. In section \ref{pseudo} the pseudocode for the directional-change
count is provided in algorithm \ref{alg:dc}.

%

\begin{figure}
\centering
\includegraphics[width=0.99\textwidth]{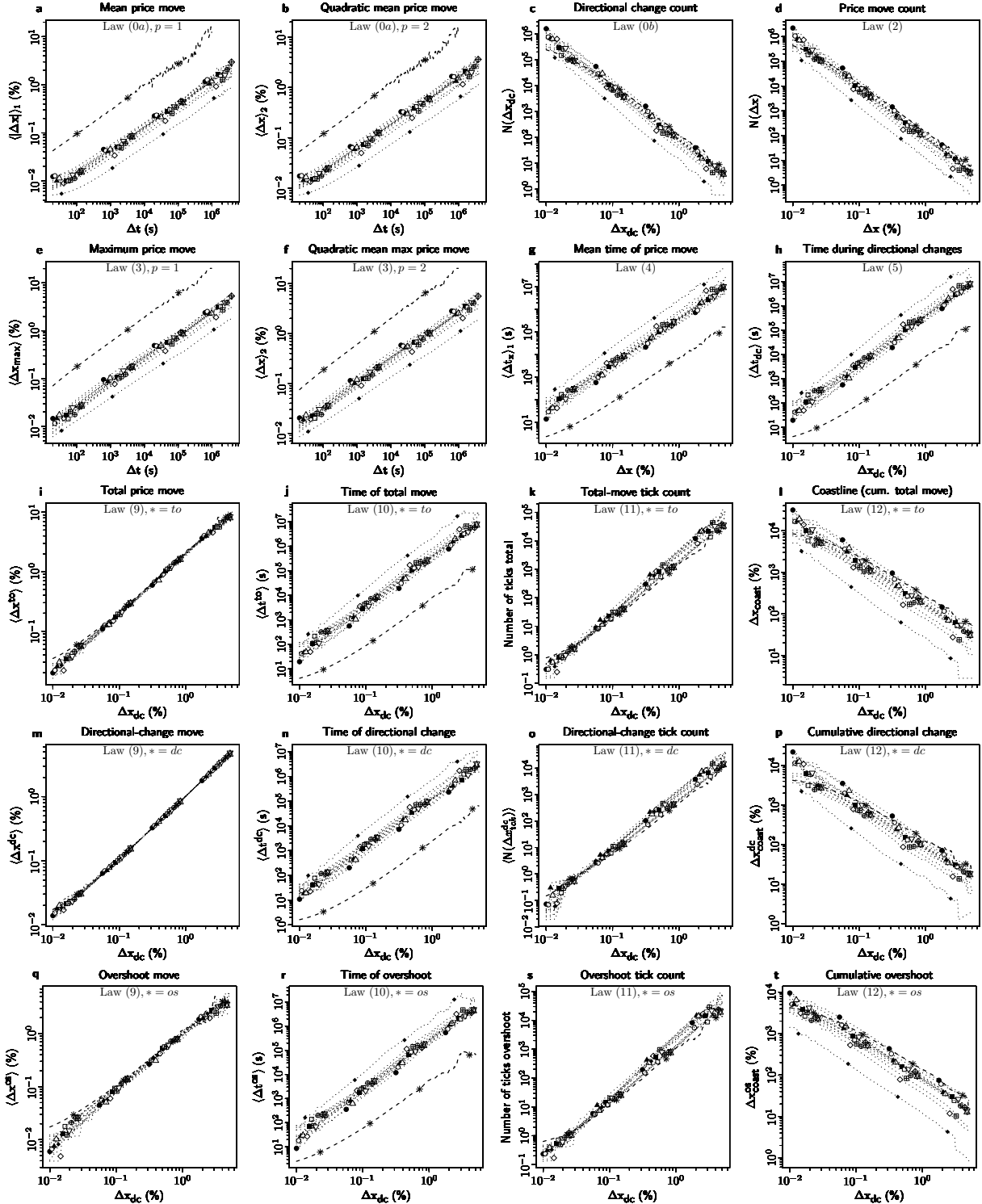}
\caption{Plots of all scaling laws described
in the text. Symbols are as in figure \ref{fig:slticks}. The raw data is
plotted for the 13 currency pairs with dots and for the  Gaussian random
walk model with dashes. See appendix \ref{app} for the values of the estimated 
scaling-law parameters.}
\label{fig:sl}
\end{figure}

Here we confirm laws (\ref{sldxdt}) and (\ref{slndx}) considering
$x_i$ (see figures \ref{fig:sl}a -- c), and report on 12 new
independent scaling laws holding across 13 exchange rates and for
close to three orders of magnitude. Appendix \ref{app} provides tables
of the estimated parameter values for all the laws and for the 13
exchange rates as well as for a Gaussian random walk (GRW) model,
described in section \ref{data}. In addition, every table lists the
average parameter values over all 13 currency pairs and their sample
standard deviations. Table \ref{t.lbl} shows the estimated scaling-law
parameters for EUR-USD. We start the enumeration of the laws by a
generalisation of equation (\ref{slndx}) that relates the average
number of ticks observed during a price move of $\Delta x$ to the size
of this threshold
\begin{equation}
\label{slticks}
\langle \textsf{N} (\Delta x_{tck}) \rangle  = \left( \frac{\Delta x}
   {C_{\textsf{\tiny N},tck}} \right)^{E_{\textsf{\tiny N},tck}},
\end{equation}
where a tick is defined as a price move larger than (in absolute
value) $\Delta x_{tck}=0.02\%$. The definition of a tick can, however,
be altered without destroying the scaling-law relation.  In essence,
this law counts the average number of ticks observed during every
price move $\Delta x$.  Law (\ref{slticks}) is plotted in figure
\ref{fig:slticks}. The second law counts the average yearly number $\textsf{N}
(\Delta x)$ of price moves of size $\Delta x$
\begin{equation}
\label{sldx}
\textsf{N} (\Delta x) = \left( \frac{ \Delta x}
    { C_{\textsf{\tiny N},x} } \right)^{E_{\textsf{\tiny N},x}}.
\end{equation}
The computation of this law is provided in algorithm \ref{alg:dx}. We
annualise the number of observations of laws (\ref{slndx}) and
(\ref{sldx}) by dividing them by 5, the number of years in our data
sample. Law (\ref{sldx}) and all the following scaling laws are given
in figure \ref{fig:sl}.
%
%
The next scaling law relates the average maximal price range $\Delta
x_{max}$, defined as the difference between the high and low price
levels, during a time interval $\Delta t$, to the size of that time
interval
\begin{equation}
\label{max}
\langle \Delta x_{max} \rangle_p  = \left( \frac{\Delta t}{C_{max}(p)} 
        \right)^{E_{max}(p)},
\end{equation}
where $\Delta x_{max}  = \max \{ x(\tau); \tau \in [t-\Delta t; t] 
\} - \min \{ x(\tau); \tau \in [t-\Delta t; t]  \}$ and
(\ref{max}) holds for $p=1,2$.
%

We have also discovered laws relating the time during which events
happen to the magnitude of these events. Law (\ref{slmdt}) relates the
average time interval $\langle \Delta t_x \rangle$ for a price change
of size $\Delta x$ to occur to the size of the threshold
\begin{equation}
\label{slmdt}
\langle \Delta t_x \rangle = \left( \frac{\Delta x}{C_{t,x}} \right)^{E_{t,x}},
\end{equation}
and similarly, considering directional changes of threshold $\Delta
x_{dc}$
\begin{equation}
\label{slmtdc}
\langle \Delta t_{dc} \rangle = \left( \frac{\Delta x_{dc}}{C_{t,dc}}
\right)^{E_{t,dc}}.
\end{equation}
Thus laws (\ref{slmdt}) and (\ref{slmtdc}) relate the average numbers
of seconds that elapse between consecutive price moves or directional
changes, respectively.


Next we unveil a set of scaling laws emerging from the identification of
directional-change events (see figure
\ref{fig:dc_to} and algorithm \ref{alg:dc}) that make up the so-called
total-move (TM) segments, which themselves decompose into
directional-change (DC) and overshoot (OS) parts. The total price
move, waiting time, and number of ticks can then be written as
\begin{eqnarray}
\langle | \Delta x^{tm} |\rangle  &=& \langle | \Delta x^{dc} |\rangle + \langle | \Delta x^{os} |\rangle , \\
\langle \Delta t^{tm} \rangle  &=& \langle  \Delta t^{dc} \rangle + \langle \Delta t^{os} \rangle,  \\
\langle  \textsf{N}( \Delta x^{tm}_{tck}) \rangle & =& \langle  \textsf{N}( \Delta x^{dc}_{tck})  \rangle + 
\langle   \textsf{N} (\Delta x^{os}_{tck})\rangle .
\end{eqnarray}
This decomposition leads to nine additional scaling laws, where the average
values are functions of the directional-change thresholds $\Delta
x_{dc}$
\begin{eqnarray}
\label{slcondx1}
\langle | \Delta x^*  |\rangle &=& \left( \frac{{\Delta x}_{dc}}{C_{x,*}} \right)^{E_{x,*}}, \\
\label{slcondx2}
\langle  \Delta t^*  \rangle  &=& \left( \frac{{\Delta x}_{dc}}{C_{t,*}} \right)^{E_{t,*}}, \\
\label{slcondx3}
\langle \textsf{N}( \Delta x^{*}_{tck}) \rangle &=& \left( \frac{{\Delta x}_{dc}}{C_{\textsf{\tiny N},*}} \right)^{E_{\textsf{\tiny N},*}},
\end{eqnarray}
where $*$ stands for $\{tm, dc, os\}$. Note that $\langle | \Delta
x^{dc} |\rangle = \Delta x_{dc}$ holds by construction. The actual
deviation to $E_{x,dc} = 1$ and $C_{x,dc} = 1$, as seen in table
\ref{tbl:11}, is given by the increasing noise for small thresholds,
as the impact of the effect of a tick exceeding the exact threshold
systematically overestimates $\langle | \Delta x^{dc} |\rangle$ (see
figure  \ref{fig:sl}m). The average parameter values (across the 13 currency pairs
given in appendix \ref{app}) of law (\ref{slcondx1}) display a
peculiar feature: on average, a directional change $\Delta x_{dc}$ is
followed by an overshoot of the same magnitude $\langle | \Delta
x^{os} | \rangle \approx \Delta x_{dc}$ ($E_{x,os}^{av} \approx 1.04$
and $C_{x,os}^{av} \approx 1.06$), making the total move double the
size of the directional-change threshold $\langle | \Delta x^{tm} |
\rangle \approx 2 \Delta x_{dc}$ ($E_{x,to}^{av} \approx 0.99$ and
$C_{x,to}^{av} \approx 0.51$).  This result is also found by computing
the probable path of the price within a binomial tree as $0.5 \;
\Delta x + 0.5^2 \; 2 \Delta x + 0.5^3 \; 3 \Delta x +
\ldots = \Delta x \sum_i^n i \;   0.5^i 
\xrightarrow{n \rightarrow \infty} 2 \Delta x$. A similar feature 
holds for the waiting times and number of ticks: $\langle | \Delta
t^{os} | \rangle \approx 2 \langle | \Delta t^{dc} |
\rangle$ and $\langle \textsf{N} ( \Delta x^{os}_{tck}) \rangle
\approx 2 \langle \textsf{N} ( \Delta x^{dc}_{tck}) \rangle$. So although
in terms of size the overshoot price move is approximately as big as
the direction-change threshold, it contains roughly twice as many ticks
and takes twice as long to unfold.

Considering cumulative price moves instead of the averages in laws
(\ref{slcondx1}) leads to another triplet of laws
\begin{equation}
\label{slcoast}
 \Delta x_{cum}^*  =  \sum_{i=1}^n  | \Delta x_i^{*}  | = \left( \frac{{\Delta x}_{dc}}{C_{cum,*}} \right)^{E_{cum,*}}.
\end{equation}
This concludes the presentation of 17 new scaling laws: we count
equation (\ref{max}) twice for $p=1,2$, and omit the trivial scaling
law $\langle | \Delta x^{dc} |\rangle \propto \Delta x_{dc}$. In the
next section, we will address the question of how many of these laws
are independent and hence can be understood as primary laws.
Our results show that most of the currency pairs exhibit similar
average behaviour. This however does not appear to be true in most of
the laws for EUR-CHF as seen in figure \ref{fig:sl}. 

It is well-known that the statistical properties of a GRW
are different from the ones observed in empirical data;
\cite{mandelbrot:04}. However, it is striking to observe how close
this simple model can be to the average properties of the market
data. Notable differences are seen in law (\ref{max}) (see figures
\ref{fig:sl}e and f) which reveals an unintuitive result: the
bell-curve distribution of price moves leads to an average maximal
price move that is roughly eight times larger than observed for the
empirical data. It should also be noted that any realisation of a GRW
has an additional degree of freedom next to the specification of the
price-move distribution $\Delta x_i \sim \mathcal{N} (\mu, \sigma^2)$,
namely the distribution of the time intervals $\Delta t$. We employ
unitary intervals $\Delta t$ of one second (see \ref{data}). It could
therefore be argued, that the additional deviations between the
empirical data and the GRW (seen in figures \ref{fig:sl}a, b, g, h, j,
n, and r) are an artifact of this simplistic choice, as they are
related to laws sensitive to the time-spacing of ticks.

 It is often claimed that scaling-laws with exponent $1/2$ are
 expected for a GRW (\cite{guillaume:97}) and hence observing this in
 empirical data would comply with the efficient market hypothesis
 (\cite{fama1970efficient}). It should be noted however, that this is
 only the case for law (\ref{sldxdt}) with $p=2$, where the exponent
 of $1/2$ can be analytically derived for a GRW. The value we actually
 measure is indeed $0.500$ (see table \ref{tbl:5}). Yet for this law
 the real market data show an exponent averaging around $0.457$.
 Furthermore, it is incorrect to assume an exponent of $1/2$ (or $2$)
 for the other laws. Indeed, our realisations of a GRW show exponents
 for all the laws but (\ref{sldxdt}) for $p=2$ to actually differ from
 $1/2$ (and $2$).

\subsection{Exploring the space of scaling laws}
\label{rel}

The scaling laws do not represent isolated patterns
but are in fact related to each other. We show that not only consistency requirements 
link the various laws, but that they can be combined to yield new scaling-law
relations.
\begin{table}[t!]
\caption{Estimated scaling law parameter values considering EUR-USD.}
\label{t.lbl}
\centering \vspace{1ex}
\begin{tabular}{lcccc}
\hline \hline 
Name & Equation & Table &  $E$  & $C$  \\ 
\hline
Tick count & \ref{slticks} & \ref{tbl:2} & $1.93$ & $2.1 \cdot  10^{-2}$ \\
Price move count &  \ref{sldx} & \ref{tbl:3} & $-1.93$ & $9.5 \cdot  10^{0}$ \\
Maximum price move &  \ref{max} ($p=1$) & \ref{tbl:6} & $0.52$ & $1.9 \cdot  10^{5}$ \\
Maximum price move &   \ref{max} ($p=2$) &  \ref{tbl:7} & $0.49$ & $1.3 \cdot  10^{5}$ \\
Time of price move &   \ref{slmdt} &  \ref{tbl:8} &  $1.93$ & $1.2 \cdot  10^{-3}$ \\
Time of directional change &   \ref{slmtdc} &  \ref{tbl:9} & $1.88$ & $1.1 \cdot  10^{-3}$ \\
Total price move &   \ref{slcondx1} &  \ref{tbl:10} & $0.98$ & $4.9 \cdot  10^{-1}$ \\
Overshoot move &   \ref{slcondx1} &  \ref{tbl:12} & $1.0$ & $9.9 \cdot  10^{-1}$ \\
Time of total move &   \ref{slcondx2} &  \ref{tbl:13} & $1.89$ & $1.1 \cdot  10^{-3}$ \\
Time of directional change &   \ref{slcondx2} &  \ref{tbl:14} & $1.85$ & $1.6 \cdot  10^{-3}$ \\
Time of overshoot &   \ref{slcondx2} &  \ref{tbl:15} & $1.91$ & $1.4 \cdot  10^{-3}$ \\
Total-move tick count &   \ref{slcondx3} &  \ref{tbl:16} & $1.89$ & $1.9 \cdot  10^{-2}$ \\
Directional-change tick count &   \ref{slcondx3} &  \ref{tbl:17} & $2.02$ & $4.2 \cdot  10^{-2}$ \\
Overshoot tick count &   \ref{slcondx3} & \ref{tbl:18}  & $1.87$ & $2.3 \cdot  10^{-2}$ \\
Cumul. total move &   \ref{slcoast} &  \ref{tbl:19} & $-0.94$ & $2.0 \cdot  10^{2}$ \\
Cumul.total move w. costs &   \ref{slcoast} &  \ref{tbl:20} & $-0.98$ & $1.5 \cdot  10^{2}$ \\
Cumul. directional change &   \ref{slcoast} & \ref{tbl:21}  & $-0.95$ & $8.8 \cdot  10^{1}$ \\
Cumul. overshoot &   \ref{slcoast} &  \ref{tbl:22} & $-0.92$ & $1.1 \cdot  10^{2}$ \\
\hline
\end{tabular}
\end{table}
To cross-check the laws we first compare laws (\ref{sldx}) and
(\ref{slmdt}), and use the fact that the average
price-move time equals the sample length divided by the number of
observations
\begin{equation}
\label{int1}
\langle \Delta t_x \rangle = Y /\mathsf{N}(\Delta x), 
\end{equation}
where $Y$ is the number of seconds in a year. This implies that 
\begin{equation}
\label{eq:crossCheck}
E_{t,x} \leftrightarrow -E_{\textsf{\tiny
N},x} \quad \textrm{and} \quad 
C_{t,x} \leftrightarrow Y^{1/E_{\textsf{\tiny
N},x}} C_{\textsf{\tiny N},x}.
\end{equation}
The estimated EUR-USD parameters from table \ref{t.lbl}  allow us to verify
equation (\ref{eq:crossCheck}) as $E_{t,x}^{} = 1.93 =
-E_{\textsf{\tiny N},x}^{}$ and $C_{t,x}^{} = 
1.23 \cdot 10^{-3}= Y^{1/E_{\textsf{\tiny N},x}^{}} 
C_{\textsf{\tiny N},x}^{}$,
where $Y = 31'553'280$ seconds. Similarly, equivalent relations hold
between laws (\ref{slndx}) and (\ref{slmtdc}): $E_{t,dc}^{} = 1.88 \approx 1.91 =
-E_{\textsf{\tiny N},dc}^{}$ and $C_{t,dc}^{} = 1.05 \cdot 10^{-3} \approx
1.11 \cdot 10^{-3}= Y^{1/E_{\textsf{\tiny N},dc}^{}} C_{\textsf{\tiny N},dc}^{}$.

Furthermore,  laws (\ref{sldxdt}) for $p=1$ and (\ref{slmdt}) are inverse relations of
each other, implying
\begin{equation}
\label{eq:crossCheck2}
E_{t,x} \leftrightarrow -E_{x}^{-1}(1) \quad \textrm{and} \quad 
C_{t,x} \leftrightarrow C_x(1)^{-E_x(1)}.
\end{equation}
We indeed have $E_{t,x}^{} = 1.93 \approx  2.01= 1 / E_x^{}(1)$ 
and $C_{t,x}^{} = 1.23  \cdot 10^{-3}   \approx  1.28 
\cdot  10^{-3}  = C_x^{}(1)^{-E_x^{}(1)} $.
There is a similar relationship to eq. (\ref{int1}) that must hold
for the cumulative and dissected average moves
\begin{equation}
\label{int2}
\langle | \Delta x^{*} | \rangle  = \Delta x_{cum}^{*} / \textsf{N} (\Delta x_{dc}),
\end{equation} 
where the asterisk stands for $\{tm, dc, os\}$.  We find for the total
move, considering EUR-USD, with a threshold of $\Delta x_{dc} = 0.1
\%$: $\langle | \Delta x^{tm} | \rangle = 0.2132 \approx 0.2129 = 1,244.2 / 5,843.4
= \Delta x_{cum}^{tm} / \textsf{N} (\Delta x_{dc})$. Similarly,
considering the directional-change and the overshoot part results in
errors smaller than $0.5\%$. In addition, $\Delta x_{cum}^{tm} =
\Delta x_{cum}^{dc} + \Delta x_{cum}^{os}$ agrees to within an error of $0.4\%$.  The
discrepancies seen in the consistency checks are in line with the
fitting errors (see appendix \ref{app}) and give us confidence to
further proceed in exploring the space of scaling laws.

The scaling laws can be assembled to produce additional laws. 
As an example, laws (\ref{sldxdt}) and (\ref{slticks}) can be used 
to relate the average number of ticks to a time interval $\Delta t$
\begin{equation}
\label{slttck}
\langle \textsf{N} (\Delta x_{tck}) \rangle =  \left( \frac{\Delta t}{C_x(1)C_{\textsf{\tiny N},tck}^{1/E_x(1)}}
\right)^{E_x(1)E_{\textsf{\tiny N},tck}}  = \left(  \frac{\Delta t}{C_{t,tck}}\right)^{E_{t,tck}},
\end{equation}
where the empirical values lead to $E_{t,tck}^{} = 0.96$, and
$C_{t,tck}^{} = 279$. This means there is a tick to be expected every
279 seconds.  This expectation is compared with law
(\ref{slmdt}) which indicates a move of $0.02\%$ (i.e., a tick)
every 258 seconds. 

In summary, we have uncovered 18 novel empirical scaling-law
relations. They provide a general self-consistent rendering of
stochastic time series, and specifically give a glimpse of the bare
bones of the market structure. There are 12 main independent scaling
laws: (\ref{slticks}), (\ref{sldx}), (\ref{max}) ($p = 1,2$),
(\ref{slcondx1}) ($tm, os$), (\ref{slcondx2}) ($tm, dc, os$), and
(\ref{slcondx3}) ($tm, dc, os$). From these, plus laws (\ref{sldxdt})
and (\ref{slndx}), six additional ones can be derived: (\ref{slmdt}),
(\ref{slmtdc}), (\ref{slcoast}) ($tm, dc, os$), and the relation given
in equation (\ref{slttck}). Note that our classification is somewhat
arbitrary, as for example law (\ref{slcondx1}) ($tm$) could be
derived from law (\ref{slcoast}) ($tm$).

\subsection{The coastline}
\label{coast}

We now have the necessary tools in hand to come back to the
measurement of the length of the coastline. The total-move scaling law
(\ref{slcoast}) allows us to estimate its size as a function of the
resolution defined by the directional-change threshold.  Considering
thresholds of $0.01\%, 0.1\%, 1\%, 5\% $, one finds the average
lengths of the annualised coastline to be $22'509\%$, $2'046\%$,
$186\%$, $34.8\%$, respectively. So by decreasing the threshold of
resolution $500$-fold, the length of the coastline decreases by a
factor of $650$.  Similarly, looking at the GRW we find $14'361\%,
1'946\%, 264\%, 65.2\%$, respectively. The $500$-fold decrease in
resolution entails a coastline decrease by a factor of only $220$,
highlighting the fact that GRW has fewer small moves and more
middle-sized moves than the empirical price curves. Not surprisingly,
taking transaction costs into account breaks the scaling law for small
thresholds. However, it is still possible to evaluate the length of
the coastline by employing the scaling relation for the interval
$[0.1\%, 5 \%]$ and measuring it for $0.05\%$. Thus, for the
thresholds $0.05\%, 0.1\%, 1\%, 5\%$ the new average coastline lengths
are now $1,604\%, 1,463\%, 161\%, 34.5\%$. For the $0.05\%$ threshold
(which occurs on average every 15 minutes), we measure an average
daily move of $6.4\%$.  The range of these average daily coastline
lengths is from $1.8 \%$ for EUR-CHF to $9.1 \%$ for AUD-JPY.

\section{Methods and data}
\label{meth}

\subsection{The data set}
\label{data}

We use a tick-by-tick database composed of 13 currency pairs spanning
 five years, from December 1, 2002 to December 1, 2007. The following
currency pairs are considered with the total number of ticks given
in parenthesis:
AUD-JPY (15'286'858),  
AUD-USD (7'037'203),  
CHF-JPY (17'081'987),  
GBP-CHF (27'141'146),   
GBP-JPY \phantom{i} (26'423'199),    
GBP-USD (13'918'523),   
EUR-AUD (19'111'129),  
EUR-GBP (13'847'688), \phantom{i} 
EUR-CHF (9'912'921),
EUR-JPY (22'594'396),
EUR-USD (13'093'081),
USD-CHF (13'812'055), 
USD-JPY (13'507'173).
The difference in the number of ticks is due to varying liquidity and
the fact that some exchange rates are synthetically generated from two
data streams, i.e., market makers quote prices based on two feeds
creating cross-rates. As an example, GBP-JPY is derived from GBP-USD
and USD-JPY. This, however, does not mean that the GBP-JPY data is
artificial as it is also real published market data. The approximately
26.5 million ticks in the GBP-JPY cross-rate are derived from the
ticks in GBP-USD (13.9 million) and USD-JPY (13.5 million).

The data is filtered as reoccurring ticks showing the same price as
the last registered tick are omitted. As the timing of price quotes
does not usually coincide with the fixed sampling times implied by an
interval $\Delta t$, for all scaling laws proportional to a power of
$\Delta t$ we use an interpolation scheme which considers the last
quoted price.  To facilitate the reproduction of our results, we
provide the USD-JPY and EUR-USD datasets that were used in our analysis
at \verb+www.olsen.ch/more/datasets/+.

In addition to the empirical data, a simple Gaussian random walk (GRW) 
model consisting of one million ticks is considered as a benchmark
\begin{equation} 
\Delta x_i  = x(t_i+ \Delta t) - x(t_i  )  \sim \mathcal{N} ( 0, \sigma^2),
\end{equation}
where $x_0 = 1.336723$, $\Delta t = 1$ second, and $\sigma =
1/6769.6$. This setting is arbitrarily chosen so as to mimic
realistic FX behaviour.

\begin{algorithm}[t]
  \caption{priceMoveCount($x$)}  \label{alg:dx}
  \begin{algorithmic}[1]
  \REQUIRE initialise variables ($n^\uparrow = n^\downarrow = 0$,  $x^{ext} = x_0$, $\Delta x \geq 0$ fixed) 
    \IF{$(x-x^{ext}) / x^{ext} \geq \Delta x$}
      \STATE $n^\uparrow \leftarrow n^\uparrow + 1$
      \STATE $x^{ext} \leftarrow x$
    \ELSIF{$(x-x^{ext} ) / x^{ext} \leq -\Delta x$}
      \STATE $n^\downarrow \leftarrow n^\downarrow + 1$
      \STATE $x^{ext} \leftarrow x$
    \ENDIF
  \end{algorithmic}
\end{algorithm}
\begin{algorithm}[h!]
  \caption{directionalChangeCount($x$)}  \label{alg:dc}
  \begin{algorithmic}[1]
  \REQUIRE initialise variables ($n^\uparrow = n^\downarrow = 0$,  $x^{ext} = x_0$, $mode$ is up, $\Delta x_{dc} \geq 0$ fixed)
  \IF{$mode$ is down}
    \IF{$x < x^{ext}$}
      \STATE  $x^{ext} \leftarrow x$
    \ELSIF{$(x-x^{ext} ) / x^{ext} \geq \Delta x$}
      \STATE $n^\uparrow \leftarrow n^\uparrow + 1$
      \STATE $x^{ext} \leftarrow x$
      \STATE $mode \leftarrow$ up
    \ENDIF
   \ELSIF{$mode$ is up}
     \IF{$x > x^{ext}$}
      \STATE  $x^{ext} \leftarrow x$
    \ELSIF{$(x-x^{ext}) / x^{ext} \leq -\Delta x$}
      \STATE $n^\downarrow \leftarrow n^\downarrow + 1$
      \STATE $x^{ext} \leftarrow x$
      \STATE $mode \leftarrow$ down
    \ENDIF
   \ENDIF
  \end{algorithmic}
\end{algorithm}

\subsection{Pseudocode}
\label{pseudo}

Algorithm \ref{alg:dx} computes the number of prices moves $n =
n^\uparrow+n^\downarrow$ for a fixed percentage threshold $\Delta
x$. Applying this computation to an array of thresholds yields law
(\ref{sldx}).

Similarly, algorithm \ref{alg:dc} counts the number of directional
changes given a threshold $\Delta x_{dc}$. Performing this calculation
for multiple thresholds recovers law (\ref{slndx}).

Excluding law (\ref{max}), the computation of all the other new scaling
laws featured in this study rely on the detection of price moves and
directional changes.

\subsection{Data fitting}
\label{fit}

It is worth noting that we do not attempt to fit power-law
distributions to empirical data (see \cite{newman:07}).  We actually
make no claim on how the data is distributed for each predefined point
of observation.  As seen in the insets of figure \ref{fig:slticks} it
is unclear to what family of distributions they belong. Rather, we
detect scaling-law relations for the average and cumulative values of
various quantities uncovered in the empirical data.

We select $250$ measurement points for laws proportional to price
thresholds, and the range is from $0.01\%$ to $5.05\%$ in logarithmic
steps: in log-space the difference of the threshold values is always
$0.025$.  For laws depending on time intervals, there are $245$
observations, and the range is from $20$ to $3'975'783$ seconds (which
is $46$ days, $23$ minutes and $3$ seconds).  The logarithmic steps
are always $0.05$.

We assume a linear relationship between the response variable $Y$ and
the random variables $X$, or $Y = A +BX$, where $A$ and $B$ are the
unknown parameters to be estimated. The actual fitting is done using
the $R$ programming language's\footnote{\texttt{www.r-project.org}.}
linear model function. In addition to the linear model, a quadratic
model is tested as an alternative hypothesis, i.e., $Y = A +BX+CX^2$,
to detect systematic curvature in the fitted data.

The tables provided in appendix \ref{app} give the estimated
scaling-law parameters for all 13 currency pairs and a GRW model, plus
their errors. In addition, we report the adjusted $R^2$ values of the
fits.  For some of the plots a slight curvature can be discerned
(e.g., for EUR-CHF in figures \ref{fig:sl}a and b). As scaling-law
relations are identified from their linear dependence, we
systematically test if a quadratic model yields a better fit. In the
last column we check for any curvature by comparing the quadratic
model's $R^2$ value to the previously reported linear value, i.e.,
$R^2_{quad} - R^2_{lin}$.  The quadratic model yields mostly a
marginally improved fit only for the GRW data.  Finally, the last row
shows the average parameter values for the currency pairs and in
parentheses the sample standard deviations.

From the linear model, it is straightforward to retrieve a scaling law relation:
\begin{equation}
y = \left( \frac{x}{C} \right)^E,
\end{equation}
where $y = \exp Y $, $x = \exp X$, $E=B$, and $C = \exp(-A/B)$. To see how the
error of $C$ propagates, we assume $A,B \sim \mathcal{N} (\mu_{A,B}, \sigma_{A,B}^2)$
and use the approximation
\begin{equation}
\textrm{Var}[C]  \approx  \left( \left.  \frac{\partial f}{\partial A}  \right|_{\mu_A, \mu_B} 
\sigma_A \right)^2 + \left( \left. \frac{\partial f}{\partial B} \right|_{\mu_A, \mu_B} \sigma_B \right)^2,
\end{equation}
where $f(A,B) = \exp(-A/B)$.

\section{Conclusions}
\label{conc}

We have enlarged the catalogue of FX stylised facts by observing 12
independent new scaling laws, next to six secondary ones, holding for
close to three orders of magnitude and across 13 currency pairs. Our
analysis relied heavily on understanding the empirical time series as
an event-based process, instead of focusing on their stochastic
nature. Relationships amongst the scaling laws can be derived and
combinations of them yield new laws. Considering an $0.05 \%$
threshold, and taking costs into account, the coastline measures on
average $6.4 \%$ {\it per day}. This is astonishingly long, and, to
our knowledge, has not been mentioned in the literature. In contrast,
on average across all currency pairs there is a mean maximal move of
$0.60\%$ to be observed within 24 hours (law (\ref{max}),
$E^{av}_{max}(1)$, $C^{av}_{max}(1)$), and on average it takes $220$
days for a move of $6.4 \%$ to be measured (law (\ref{slmdt})).  This
indicates the importance of considering not only the tail events
associated with crisis, but also accounting for the numerous smaller
events that precede this event.

In finance, where frames of reference and fixed points are hard to come by and often illusory, 
the new scaling laws provide a reliable framework.  We believe they can enhance our study of 
the dynamic behaviour of markets and improve the quality of the inferences and predictions 
we make about the behaviour of prices.  The new laws represent the foundation of a 
completely new generation of tools for studying volatility, measuring risk, and creating 
better forecasting and trading models.

\section{Acknowledgements}

We thank L. Wilkens for helping with figure \ref{fig:dc_to}; T. Grizzard for editing
the manuscript; and the Seminar for Statistics at the ETH in Zurich for their
consultation.


\bibliography{bibfile}
\bibliographystyle{plainnat}

\newpage

\begin{appendix}

\renewcommand{\thetable}{A\arabic{table}}
\addtocounter{table}{-1}

\section{Tables}
\label{app}

\begin{table}[h]
\caption{Directional change count, law (\ref{slndx})}
\label{tbl:1}
\centering
\begin{tabular}{lcccccc} \\ [-2.5ex]
\hline  \\  [-2.1ex]
Currency & $E_{\textsf{\tiny N},dc}$ & $\Delta E_{\textsf{\tiny N},dc}$ & $ C_{\textsf{\tiny{N}},dc}$  & $\Delta  C_{\textsf{\tiny{N}},dc}$ & Adj. $R^2$&   $R^2_{quad} - R^2_{lin}$ \\   [0.5ex]
\hline \\ [-2.1ex]
AUD-JPY  &  -2.046  & $\pm$  4.6e-03  &  1.116e+01  & $\pm$ 8.3e-02  &  0.99877  &  2.775e-04 \\ 
AUD-USD  &  -1.949  & $\pm$  4.4e-03  &  1.136e+01  & $\pm$ 8.7e-02  &  0.99873  &  4.807e-04 \\ 
CHF-JPY  &  -2.067  & $\pm$  4.7e-03  &  8.699e+00  & $\pm$ 6.3e-02  &  0.99871  &  1.068e-04 \\ 
EUR-AUD  &  -2.133  & $\pm$  5.8e-03  &  8.233e+00  & $\pm$ 7.0e-02  &  0.99818  &  6.540e-05 \\ 
EUR-CHF  &  -2.158  & $\pm$  5.4e-03  &  3.218e+00  & $\pm$ 2.1e-02  &  0.99844  &  2.642e-04 \\ 
EUR-GBP  &  -2.178  & $\pm$  7.6e-03  &  5.430e+00  & $\pm$ 5.5e-02  &  0.99696  &  1.004e-03 \\ 
EUR-JPY  &  -2.002  & $\pm$  2.9e-03  &  9.083e+00  & $\pm$ 4.2e-02  &  0.99949  &  1.090e-04 \\ 
EUR-USD  &  -1.908  & $\pm$  5.0e-03  &  9.422e+00  & $\pm$ 8.1e-02  &  0.99827  &  1.031e-03 \\ 
GBP-CHF  &  -2.131  & $\pm$  3.1e-03  &  6.406e+00  & $\pm$ 2.7e-02  &  0.99949  &  4.152e-05 \\ 
GBP-JPY  &  -2.017  & $\pm$  3.4e-03  &  9.440e+00  & $\pm$ 5.1e-02  &  0.99931  &  -2.733e-06 \\ 
GBP-USD  &  -1.904  & $\pm$  3.2e-03  &  8.947e+00  & $\pm$ 4.8e-02  &  0.99931  &  2.562e-04 \\ 
GRW  &  -1.797  & $\pm$  9.3e-03  &  1.528e+01  & $\pm$ 2.8e-01  &  0.99337  &  5.137e-03 \\ 
USD-CHF  &  -1.908  & $\pm$  3.3e-03  &  1.070e+01  & $\pm$ 6.1e-02  &  0.99928  &  3.056e-04 \\ 
USD-JPY  &  -1.928  & $\pm$  4.6e-03  &  9.841e+00  & $\pm$ 7.7e-02  &  0.99857  &  7.382e-04 \\ 
\hline \\ [-2.1ex]
Currency average & -2.03 & (1.0e-01) & 8.61e+00 & (2.3e+00) & & \\
\end{tabular}
\end{table}

\begin{table}[h!]
\caption{Tick count, law (\ref{slticks}),  $\Delta x_{tck} = 0.02\%$}
\label{tbl:2}
\centering
\begin{tabular}{lcccccc} \\ [-2.1ex]
\hline  \\  [-2.1ex]
Currency & $E_{\textsf{\tiny N},tck}$ & $\Delta E_{\textsf{\tiny N},tck}$ & $ C_{\textsf{\tiny{N}},tck}$  & $\Delta  C_{\textsf{\tiny{N}},tck}$ & Adj. $R^2$&   $R^2_{quad} - R^2_{lin}$ \\   [0.5ex]
\hline \\ [-2.1ex]
AUD-JPY  &  2.051  & $\pm$  4.0e-03  &  1.879e-02  & $\pm$ 1.7e-04  &  0.99906  &  4.957e-05 \\ 
AUD-USD  &  1.970  & $\pm$  3.8e-03  &  2.146e-02  & $\pm$ 1.9e-04  &  0.99906  &  -3.722e-06 \\ 
CHF-JPY  &  2.085  & $\pm$  5.4e-03  &  1.663e-02  & $\pm$ 2.0e-04  &  0.99833  &  1.392e-04 \\ 
EUR-AUD  &  2.134  & $\pm$  4.2e-03  &  2.028e-02  & $\pm$ 1.8e-04  &  0.99904  &  1.879e-04 \\ 
EUR-CHF  &  2.120  & $\pm$  5.2e-03  &  2.053e-02  & $\pm$ 2.3e-04  &  0.99848  &  7.283e-04 \\ 
EUR-GBP  &  2.185  & $\pm$  9.4e-03  &  2.150e-02  & $\pm$ 4.2e-04  &  0.99541  &  1.699e-03 \\ 
EUR-JPY  &  1.997  & $\pm$  3.0e-03  &  1.864e-02  & $\pm$ 1.3e-04  &  0.99946  &  1.960e-04 \\ 
EUR-USD  &  1.928  & $\pm$  3.2e-03  &  2.099e-02  & $\pm$ 1.6e-04  &  0.99933  &  1.827e-04 \\ 
GBP-CHF  &  2.122  & $\pm$  3.0e-03  &  1.931e-02  & $\pm$ 1.3e-04  &  0.99950  &  1.265e-04 \\ 
GBP-JPY  &  2.027  & $\pm$  2.7e-03  &  1.920e-02  & $\pm$ 1.2e-04  &  0.99955  &  -1.793e-06 \\ 
GBP-USD  &  1.932  & $\pm$  2.2e-03  &  2.035e-02  & $\pm$ 1.1e-04  &  0.99967  &  6.484e-05 \\ 
GRW  &  1.864  & $\pm$  6.0e-03  &  2.112e-02  & $\pm$ 3.1e-04  &  0.99740  &  1.838e-03 \\ 
USD-CHF  &  1.945  & $\pm$  3.2e-03  &  2.027e-02  & $\pm$ 1.5e-04  &  0.99932  &  2.199e-04 \\ 
USD-JPY  &  1.975  & $\pm$  3.9e-03  &  2.206e-02  & $\pm$ 2.0e-04  &  0.99901  &  4.497e-04 \\ 
\hline \\ [-2.1ex]
Currency average & 2.04 & (8.6e-02) & 2.0e-02 & (1.5e-03) & & \\
\end{tabular}
\end{table}

\clearpage

\begin{table}[h!]
\caption{Price move count, law (\ref{sldx})}
\label{tbl:3}
\centering
\begin{tabular}{lcccccc} \\ [-2.1ex]
\hline  \\  [-2.1ex]
Currency & $E_{\textsf{\tiny N},x}$ & $\Delta E_{\textsf{\tiny N},x}$ & $ C_{\textsf{\tiny{N}},x}$  & $\Delta  C_{\textsf{\tiny{N}},x}$ & Adj. $R^2$&   $R^2_{quad} - R^2_{lin}$ \\   [0.5ex]
\hline \\ [-2.1ex]
AUD-JPY  &  -2.051  & $\pm$  4.0e-03  &  1.111e+01  & $\pm$ 7.2e-02  &  0.99907  &  4.533e-05 \\ 
AUD-USD  &  -1.973  & $\pm$  3.8e-03  &  1.125e+01  & $\pm$ 7.3e-02  &  0.99907  &  -2.901e-06 \\ 
CHF-JPY  &  -2.092  & $\pm$  5.4e-03  &  8.095e+00  & $\pm$ 6.5e-02  &  0.99837  &  8.115e-05 \\ 
EUR-AUD  &  -2.135  & $\pm$  4.1e-03  &  7.912e+00  & $\pm$ 4.8e-02  &  0.99907  &  1.727e-04 \\ 
EUR-CHF  &  -2.138  & $\pm$  4.8e-03  &  3.171e+00  & $\pm$ 1.9e-02  &  0.99875  &  3.725e-04 \\ 
EUR-GBP  &  -2.187  & $\pm$  9.3e-03  &  5.141e+00  & $\pm$ 6.3e-02  &  0.99547  &  1.638e-03 \\ 
EUR-JPY  &  -2.001  & $\pm$  2.8e-03  &  9.137e+00  & $\pm$ 4.1e-02  &  0.99951  &  1.469e-04 \\ 
EUR-USD  &  -1.930  & $\pm$  3.2e-03  &  9.469e+00  & $\pm$ 5.1e-02  &  0.99932  &  2.033e-04 \\ 
GBP-CHF  &  -2.124  & $\pm$  3.0e-03  &  6.210e+00  & $\pm$ 2.6e-02  &  0.99951  &  1.162e-04 \\ 
GBP-JPY  &  -2.029  & $\pm$  2.8e-03  &  9.142e+00  & $\pm$ 4.1e-02  &  0.99953  &  -3.668e-07 \\ 
GBP-USD  &  -1.936  & $\pm$  2.3e-03  &  8.758e+00  & $\pm$ 3.3e-02  &  0.99965  &  9.540e-05 \\ 
GRW  &  -1.866  & $\pm$  6.1e-03  &  1.419e+01  & $\pm$ 1.6e-01  &  0.99737  &  1.907e-03 \\ 
USD-CHF  &  -1.946  & $\pm$  3.2e-03  &  1.022e+01  & $\pm$ 5.6e-02  &  0.99931  &  2.329e-04 \\ 
USD-JPY  &  -1.978  & $\pm$  4.0e-03  &  9.048e+00  & $\pm$ 5.9e-02  &  0.99897  &  4.988e-04 \\ 
\hline \\ [-2.1ex]
Currency average & -2.04 & (8.8e-02) & 8.36e+00 & (2.3e+00) & & \\
\end{tabular}
\end{table}

\begin{table}[h!]
\caption{Mean price move during $\Delta t$, law (\ref{sldxdt}), $p=1$}
\label{tbl:4}
\centering
\begin{tabular}{lcccccc} \\ [-2.1ex]
\hline  \\  [-2.1ex]
Currency & $E_{x}(1)$ & $\Delta E_{x}(1)$ & $ C_{x}(1)$  & $\Delta  C_{x}(1)$ & Adj. $R^2_{lin}$&   $R^2_{quad} - R^2_{lin}$ \\   [0.5ex]
\hline \\ [-2.1ex]
AUD-JPY  &  0.462  & $\pm$  1.3e-03  &  4.783e+05  & $\pm$ 2.2e+04  &  0.99809  &  1.261e-03 \\ 
AUD-USD  &  0.473  & $\pm$  1.8e-03  &  4.779e+05  & $\pm$ 3.0e+04  &  0.99646  &  2.108e-03 \\ 
CHF-JPY  &  0.461  & $\pm$  1.1e-03  &  7.962e+05  & $\pm$ 3.1e+04  &  0.99872  &  6.185e-05 \\ 
EUR-AUD  &  0.451  & $\pm$  1.3e-03  &  8.584e+05  & $\pm$ 4.1e+04  &  0.99802  &  1.233e-03 \\ 
EUR-CHF  &  0.450  & $\pm$  2.0e-03  &  6.362e+06  & $\pm$ 5.1e+05  &  0.99538  &  3.115e-03 \\ 
EUR-GBP  &  0.456  & $\pm$  1.8e-03  &  1.825e+06  & $\pm$ 1.2e+05  &  0.99630  &  2.198e-03 \\ 
EUR-JPY  &  0.483  & $\pm$  7.6e-04  &  6.899e+05  & $\pm$ 1.8e+04  &  0.99940  &  1.606e-04 \\ 
EUR-USD  &  0.497  & $\pm$  1.1e-03  &  6.632e+05  & $\pm$ 2.5e+04  &  0.99875  &  7.131e-04 \\ 
GBP-CHF  &  0.462  & $\pm$  8.8e-04  &  1.317e+06  & $\pm$ 4.3e+04  &  0.99913  &  3.825e-04 \\ 
GBP-JPY  &  0.476  & $\pm$  7.7e-04  &  6.637e+05  & $\pm$ 1.8e+04  &  0.99936  &  2.521e-04 \\ 
GBP-USD  &  0.496  & $\pm$  9.9e-04  &  7.517e+05  & $\pm$ 2.5e+04  &  0.99903  &  5.506e-04 \\ 
GRW  &  0.510  & $\pm$  3.1e-03  &  1.070e+04  & $\pm$ 8.4e+02  &  0.99200  &  8.416e-04 \\ 
USD-CHF  &  0.493  & $\pm$  9.6e-04  &  5.516e+05  & $\pm$ 1.8e+04  &  0.99907  &  2.802e-04 \\ 
USD-JPY  &  0.482  & $\pm$  9.3e-04  &  6.949e+05  & $\pm$ 2.2e+04  &  0.99909  &  3.420e-04 \\ 
\hline \\ [-2.1ex]
Currency average & 0.47 & (1.7e-02) & 1.24e+06 & (1.6e+06) & & \\
\end{tabular}
\end{table}

\clearpage

\begin{table}[h!]
\caption{Quadratic mean price move (historical volatility) during $\Delta t$, law (\ref{sldxdt}), $p=2$}
\label{tbl:5}
\centering
\begin{tabular}{lcccccc} \\ [-2.1ex]
\hline  \\  [-2.1ex]
Currency &  $E_{x}(2)$ & $\Delta E_{x}(2)$ & $ C_{x}(2)$  & $\Delta  C_{x}(2)$ & Adj. $R^2_{lin}$&   $R^2_{quad} - R^2_{lin}$ \\   [0.5ex]
\hline \\ [-2.1ex]
AUD-JPY  &  0.454  & $\pm$  9.5e-04  &  2.274e+05  & $\pm$ 7.5e+03  &  0.99892  &  3.195e-04 \\ 
AUD-USD  &  0.458  & $\pm$  1.2e-03  &  2.616e+05  & $\pm$ 1.1e+04  &  0.99822  &  8.464e-04 \\ 
CHF-JPY  &  0.449  & $\pm$  8.6e-04  &  4.225e+05  & $\pm$ 1.3e+04  &  0.99910  &  -2.042e-06 \\ 
EUR-AUD  &  0.441  & $\pm$  1.1e-03  &  4.467e+05  & $\pm$ 1.8e+04  &  0.99852  &  8.995e-04 \\ 
EUR-CHF  &  0.430  & $\pm$  1.4e-03  &  3.888e+06  & $\pm$ 2.3e+05  &  0.99727  &  1.685e-03 \\ 
EUR-GBP  &  0.440  & $\pm$  1.5e-03  &  1.018e+06  & $\pm$ 5.7e+04  &  0.99730  &  1.473e-03 \\ 
EUR-JPY  &  0.467  & $\pm$  6.3e-04  &  3.852e+05  & $\pm$ 8.4e+03  &  0.99955  &  -1.412e-06 \\ 
EUR-USD  &  0.473  & $\pm$  7.6e-04  &  3.843e+05  & $\pm$ 1.0e+04  &  0.99936  &  6.888e-05 \\ 
GBP-CHF  &  0.450  & $\pm$  6.3e-04  &  7.212e+05  & $\pm$ 1.7e+04  &  0.99951  &  7.869e-05 \\ 
GBP-JPY  &  0.463  & $\pm$  6.2e-04  &  3.588e+05  & $\pm$ 7.8e+03  &  0.99956  &  5.572e-06 \\ 
GBP-USD  &  0.475  & $\pm$  6.8e-04  &  4.399e+05  & $\pm$ 1.0e+04  &  0.99950  &  9.130e-05 \\ 
GRW  &  0.500  & $\pm$  2.3e-03  &  7.133e+03  & $\pm$ 4.1e+02  &  0.99548  &  1.762e-05 \\ 
USD-CHF  &  0.472  & $\pm$  7.6e-04  &  3.131e+05  & $\pm$ 8.1e+03  &  0.99937  &  -1.410e-06 \\ 
USD-JPY  &  0.463  & $\pm$  6.5e-04  &  3.912e+05  & $\pm$ 8.9e+03  &  0.99952  &  3.957e-05 \\ 
\hline \\ [-2.1ex]
Currency average & 0.46 & (1.4e-02) & 7.12e+05 & (9.8e+05) & & \\
\end{tabular}
\end{table}

\begin{table}[h!]
\caption{Maximal price move during $\Delta t$, law (\ref{max}), $p=1$}
\label{tbl:6}
\centering
\begin{tabular}{lcccccc} \\ [-2.1ex]
\hline  \\  [-2.1ex]
Currency & $E_{max}(1)$ & $\Delta E_{max}(1)$ & $ C_{max}(1)$  & $\Delta  C_{max}(1)$ & Adj. $R^2$&   $R^2_{quad} - R^2_{lin}$ \\   [0.5ex]
\hline \\ [-2.1ex]
AUD-JPY  &  0.479  & $\pm$  1.1e-03  &  9.743e+04  & $\pm$ 3.4e+03  &  0.99867  &  5.190e-04 \\ 
AUD-USD  &  0.511  & $\pm$  1.2e-03  &  1.106e+05  & $\pm$ 3.9e+03  &  0.99867  &  8.429e-04 \\ 
CHF-JPY  &  0.478  & $\pm$  7.1e-04  &  1.515e+05  & $\pm$ 3.4e+03  &  0.99947  &  3.451e-04 \\ 
EUR-AUD  &  0.464  & $\pm$  1.1e-03  &  1.523e+05  & $\pm$ 5.4e+03  &  0.99872  &  3.016e-04 \\ 
EUR-CHF  &  0.466  & $\pm$  4.2e-04  &  1.084e+06  & $\pm$ 1.7e+04  &  0.99980  &  7.757e-06 \\ 
EUR-GBP  &  0.467  & $\pm$  4.6e-04  &  3.410e+05  & $\pm$ 5.4e+03  &  0.99977  &  4.654e-06 \\ 
EUR-JPY  &  0.495  & $\pm$  7.8e-04  &  1.572e+05  & $\pm$ 3.8e+03  &  0.99939  &  2.617e-04 \\ 
EUR-USD  &  0.521  & $\pm$  1.1e-03  &  1.676e+05  & $\pm$ 5.7e+03  &  0.99884  &  8.795e-04 \\ 
GBP-CHF  &  0.469  & $\pm$  7.2e-04  &  2.528e+05  & $\pm$ 6.1e+03  &  0.99942  &  1.830e-04 \\ 
GBP-JPY  &  0.487  & $\pm$  9.5e-04  &  1.426e+05  & $\pm$ 4.3e+03  &  0.99908  &  3.303e-04 \\ 
GBP-USD  &  0.522  & $\pm$  1.2e-03  &  1.873e+05  & $\pm$ 6.5e+03  &  0.99880  &  8.679e-04 \\ 
GRW  &  0.513  & $\pm$  1.0e-03  &  2.996e+03  & $\pm$ 7.2e+01  &  0.99914  &  6.057e-05 \\ 
USD-CHF  &  0.516  & $\pm$  1.1e-03  &  1.355e+05  & $\pm$ 4.4e+03  &  0.99893  &  8.976e-04 \\ 
USD-JPY  &  0.508  & $\pm$  1.2e-03  &  1.579e+05  & $\pm$ 5.7e+03  &  0.99865  &  1.020e-03 \\ 
\hline \\ [-2.1ex]
Currency average & 0.49 & (2.2e-02) & 2.41e+05 & (2.6e+05) & & \\
\end{tabular}
\end{table}

\begin{table}[h!]
\caption{Quadratic maximal price move during $\Delta t$, law (\ref{max}), $p=2$}
\label{tbl:7}
\centering
\begin{tabular}{lcccccc} \\ [-2.1ex]
\hline  \\  [-2.1ex]
Currency & $E_{max}(2)$ & $\Delta E_{max}(2)$ & $ C_{max}(2)$  & $\Delta  C_{max}(2)$ & Adj. $R^2$&   $R^2_{quad} - R^2_{lin}$ \\   [0.5ex]
\hline \\ [-2.1ex]
AUD-JPY  &  0.470  & $\pm$  7.5e-04  &  6.850e+04  & $\pm$ 1.6e+03  &  0.99939  &  2.267e-04 \\ 
AUD-USD  &  0.486  & $\pm$  7.5e-04  &  8.513e+04  & $\pm$ 2.0e+03  &  0.99942  &  2.839e-04 \\ 
CHF-JPY  &  0.466  & $\pm$  6.0e-04  &  1.198e+05  & $\pm$ 2.4e+03  &  0.99959  &  2.618e-04 \\ 
EUR-AUD  &  0.450  & $\pm$  6.7e-04  &  1.191e+05  & $\pm$ 2.7e+03  &  0.99947  &  2.459e-05 \\ 
EUR-CHF  &  0.446  & $\pm$  3.8e-04  &  9.439e+05  & $\pm$ 1.4e+04  &  0.99982  &  2.594e-06 \\ 
EUR-GBP  &  0.450  & $\pm$  4.0e-04  &  2.806e+05  & $\pm$ 4.0e+03  &  0.99980  &  3.721e-06 \\ 
EUR-JPY  &  0.481  & $\pm$  6.4e-04  &  1.229e+05  & $\pm$ 2.5e+03  &  0.99957  &  2.387e-04 \\ 
EUR-USD  &  0.494  & $\pm$  1.1e-03  &  1.337e+05  & $\pm$ 4.4e+03  &  0.99887  &  8.416e-04 \\ 
GBP-CHF  &  0.456  & $\pm$  5.6e-04  &  2.050e+05  & $\pm$ 3.9e+03  &  0.99963  &  1.333e-04 \\ 
GBP-JPY  &  0.475  & $\pm$  7.4e-04  &  1.106e+05  & $\pm$ 2.6e+03  &  0.99940  &  2.845e-04 \\ 
GBP-USD  &  0.495  & $\pm$  9.7e-04  &  1.529e+05  & $\pm$ 4.6e+03  &  0.99907  &  6.398e-04 \\ 
GRW  &  0.510  & $\pm$  1.0e-03  &  2.739e+03  & $\pm$ 6.5e+01  &  0.99914  &  1.415e-04 \\ 
USD-CHF  &  0.493  & $\pm$  1.1e-03  &  1.080e+05  & $\pm$ 3.8e+03  &  0.99871  &  1.042e-03 \\ 
USD-JPY  &  0.485  & $\pm$  1.0e-03  &  1.259e+05  & $\pm$ 4.1e+03  &  0.99888  &  8.555e-04 \\ 
\hline \\ [-2.1ex]
Currency average & 0.4728 & (1.8e-0) & 1.98e+05 & (2.3e+05) & & \\
\end{tabular}
\end{table}

\begin{table}[h!]
\caption{Time of price move, law (\ref{slmdt})}
\label{tbl:8}
\centering
\begin{tabular}{lcccccc} \\ [-2.1ex]
\hline  \\  [-2.1ex]
Currency & $E_{t,x}$ & $\Delta E_{t,x}$ & $ C_{t,x}$  & $\Delta  C_{t,x}$ & Adj. $R^2$&   $R^2_{quad} - R^2_{lin}$ \\   [0.5ex]
\hline \\ [-2.1ex]
AUD-JPY  &  2.051  & $\pm$  4.0e-03  &  2.477e-03  & $\pm$ 3.1e-05  &  0.99906  &  4.705e-05 \\ 
AUD-USD  &  1.972  & $\pm$  3.8e-03  &  1.774e-03  & $\pm$ 2.3e-05  &  0.99907  &  -3.584e-06 \\ 
CHF-JPY  &  2.088  & $\pm$  5.4e-03  &  2.109e-03  & $\pm$ 3.6e-05  &  0.99835  &  1.145e-04 \\ 
EUR-AUD  &  2.134  & $\pm$  4.2e-03  &  2.451e-03  & $\pm$ 3.1e-05  &  0.99905  &  1.803e-04 \\ 
EUR-CHF  &  2.125  & $\pm$  5.1e-03  &  9.598e-04  & $\pm$ 1.7e-05  &  0.99858  &  6.277e-04 \\ 
EUR-GBP  &  2.184  & $\pm$  9.4e-03  &  1.905e-03  & $\pm$ 5.5e-05  &  0.99538  &  1.725e-03 \\ 
EUR-JPY  &  1.999  & $\pm$  2.9e-03  &  1.627e-03  & $\pm$ 1.6e-05  &  0.99947  &  1.794e-04 \\ 
EUR-USD  &  1.928  & $\pm$  3.2e-03  &  1.227e-03  & $\pm$ 1.4e-05  &  0.99933  &  1.853e-04 \\ 
GBP-CHF  &  2.123  & $\pm$  3.0e-03  &  1.825e-03  & $\pm$ 1.7e-05  &  0.99951  &  1.237e-04 \\ 
GBP-JPY  &  2.028  & $\pm$  2.7e-03  &  1.841e-03  & $\pm$ 1.7e-05  &  0.99954  &  -1.336e-06 \\ 
GBP-USD  &  1.932  & $\pm$  2.2e-03  &  1.162e-03  & $\pm$ 9.5e-06  &  0.99967  &  6.721e-05 \\ 
GRW  &  1.864  & $\pm$  6.0e-03  &  8.640e-03  & $\pm$ 1.5e-04  &  0.99740  &  1.843e-03 \\ 
USD-CHF  &  1.945  & $\pm$  3.2e-03  &  1.428e-03  & $\pm$ 1.6e-05  &  0.99932  &  2.213e-04 \\ 
USD-JPY  &  1.977  & $\pm$  4.0e-03  &  1.461e-03  & $\pm$ 2.0e-05  &  0.99898  &  4.814e-04 \\ 
\hline \\ [-2.1ex]
Currency average & 2.04 & (8.6e-02) & 1.71e-03 & (4.7e-04) & & \\
\end{tabular}
\end{table}

\begin{table}[h!]
\caption{Time between directional changes, law (\ref{slmtdc})}
\label{tbl:9}
\centering
\begin{tabular}{lcccccc} \\ [-2.1ex]
\hline  \\  [-2.1ex]
Currency & $E_{t,dc}$ & $\Delta E_{t,dc}$ & $ C_{t,dc}$  & $\Delta  C_{t,dc}$ & Adj. $R^2$&   $R^2_{quad} - R^2_{lin}$ \\   [0.5ex]
\hline \\ [-2.1ex]
AUD-JPY  &  2.046  & $\pm$  4.5e-03  &  2.439e-03  & $\pm$ 3.5e-05  &  0.99877  &  2.749e-04 \\ 
AUD-USD  &  1.948  & $\pm$  4.4e-03  &  1.608e-03  & $\pm$ 2.5e-05  &  0.99873  &  4.641e-04 \\ 
CHF-JPY  &  2.063  & $\pm$  4.8e-03  &  2.053e-03  & $\pm$ 3.2e-05  &  0.99863  &  1.438e-04 \\ 
EUR-AUD  &  2.132  & $\pm$  5.8e-03  &  2.531e-03  & $\pm$ 4.4e-05  &  0.99817  &  6.042e-05 \\ 
EUR-CHF  &  2.111  & $\pm$  6.8e-03  &  9.819e-04  & $\pm$ 2.3e-05  &  0.99741  &  1.586e-03 \\ 
EUR-GBP  &  2.162  & $\pm$  8.3e-03  &  1.907e-03  & $\pm$ 4.9e-05  &  0.99635  &  1.536e-03 \\ 
EUR-JPY  &  2.001  & $\pm$  2.9e-03  &  1.629e-03  & $\pm$ 1.6e-05  &  0.99947  &  1.180e-04 \\ 
EUR-USD  &  1.884  & $\pm$  3.9e-03  &  1.050e-03  & $\pm$ 1.6e-05  &  0.99893  &  3.450e-04 \\ 
GBP-CHF  &  2.127  & $\pm$  3.2e-03  &  1.926e-03  & $\pm$ 1.9e-05  &  0.99944  &  7.375e-05 \\ 
GBP-JPY  &  2.016  & $\pm$  3.4e-03  &  1.804e-03  & $\pm$ 2.0e-05  &  0.99930  &  -2.835e-06 \\ 
GBP-USD  &  1.899  & $\pm$  3.2e-03  &  1.020e-03  & $\pm$ 1.3e-05  &  0.99928  &  1.643e-04 \\ 
GRW  &  1.790  & $\pm$  9.1e-03  &  6.953e-03  & $\pm$ 1.9e-04  &  0.99361  &  4.807e-03 \\ 
USD-CHF  &  1.904  & $\pm$  3.1e-03  &  1.247e-03  & $\pm$ 1.5e-05  &  0.99932  &  2.326e-04 \\ 
USD-JPY  &  1.927  & $\pm$  4.6e-03  &  1.266e-03  & $\pm$ 2.1e-05  &  0.99857  &  7.180e-04 \\ 
\hline \\ [-2.1ex]
Currency average & 2.02 & (9.8e-02) & 1.65e-03 & (5.2e-04) & & \\
\end{tabular}
\end{table}

\begin{table}
\caption{Total price move, law (\ref{slcondx1}), $*=tm$}
\label{tbl:10}
\centering
\begin{tabular}{lcccccc} \\ [-2.5ex]
\hline  \\  [-2.1ex]
Currency & $E_{x,tm}$ & $\Delta E_{x,tm}$ & $ C_{x,tm}$  & $\Delta  C_{x,tm}$ & Adj. $R^2$&   $R^2_{quad} - R^2_{lin}$ \\   [0.5ex]
\hline \\ [-2.1ex]
AUD-JPY  &  1.001  & $\pm$  2.3e-03  &  4.998e-01  & $\pm$ 2.8e-03  &  0.99872  &  4.740e-04 \\ 
AUD-USD  &  0.990  & $\pm$  2.0e-03  &  4.876e-01  & $\pm$ 2.4e-03  &  0.99899  &  2.066e-04 \\ 
CHF-JPY  &  0.995  & $\pm$  2.3e-03  &  5.211e-01  & $\pm$ 2.9e-03  &  0.99868  &  2.848e-05 \\ 
EUR-AUD  &  0.997  & $\pm$  2.4e-03  &  5.232e-01  & $\pm$ 3.1e-03  &  0.99851  &  3.814e-04 \\ 
EUR-CHF  &  1.006  & $\pm$  2.3e-03  &  5.299e-01  & $\pm$ 3.0e-03  &  0.99865  &  2.258e-04 \\ 
EUR-GBP  &  1.004  & $\pm$  3.0e-03  &  5.388e-01  & $\pm$ 3.9e-03  &  0.99782  &  2.652e-04 \\ 
EUR-JPY  &  1.001  & $\pm$  1.3e-03  &  5.001e-01  & $\pm$ 1.6e-03  &  0.99956  &  6.695e-06 \\ 
EUR-USD  &  0.976  & $\pm$  1.7e-03  &  4.871e-01  & $\pm$ 2.1e-03  &  0.99921  &  2.324e-04 \\ 
GBP-CHF  &  0.993  & $\pm$  1.6e-03  &  5.358e-01  & $\pm$ 2.2e-03  &  0.99932  &  4.097e-05 \\ 
GBP-JPY  &  0.996  & $\pm$  1.5e-03  &  5.039e-01  & $\pm$ 1.9e-03  &  0.99940  &  2.969e-05 \\ 
GBP-USD  &  0.981  & $\pm$  1.2e-03  &  4.885e-01  & $\pm$ 1.5e-03  &  0.99963  &  3.725e-05 \\ 
GRW  &  0.943  & $\pm$  3.4e-03  &  4.708e-01  & $\pm$ 4.2e-03  &  0.99670  &  2.060e-03 \\ 
USD-CHF  &  0.973  & $\pm$  1.4e-03  &  4.984e-01  & $\pm$ 1.7e-03  &  0.99950  &  -1.570e-06 \\ 
USD-JPY  &  0.969  & $\pm$  1.8e-03  &  5.028e-01  & $\pm$ 2.3e-03  &  0.99913  &  1.081e-04 \\ 
\hline \\ [-2.1ex]
Currency average & 0.99 & (1.2e-02) & 5.10e-01 & (1.9e-02) & & \\
\end{tabular}
\end{table}

\begin{table}
\caption{Directional-change move, law (\ref{slcondx1}), $*=dc$}
\label{tbl:11}
\centering
\begin{tabular}{lcccccc} \\ [-2.5ex]
\hline  \\  [-2.1ex]
Currency & $E_{x,dc}$ & $\Delta E_{x,dc}$ & $ C_{x,dc}$  & $\Delta  C_{x,dc}$ & Adj. $R^2$&   $R^2_{quad} - R^2_{lin}$ \\   [0.5ex]
\hline \\ [-2.1ex]
AUD-JPY  &  0.941  & $\pm$  2.5e-03  &  9.847e-01  & $\pm$ 6.1e-03  &  0.99828  &  1.033e-03 \\ 
AUD-USD  &  0.937  & $\pm$  2.4e-03  &  9.814e-01  & $\pm$ 5.9e-03  &  0.99838  &  9.673e-04 \\ 
CHF-JPY  &  0.949  & $\pm$  2.6e-03  &  9.898e-01  & $\pm$ 6.3e-03  &  0.99816  &  8.473e-04 \\ 
EUR-AUD  &  0.946  & $\pm$  2.0e-03  &  9.843e-01  & $\pm$ 5.0e-03  &  0.99884  &  8.851e-04 \\ 
EUR-CHF  &  0.964  & $\pm$  1.5e-03  &  9.907e-01  & $\pm$ 3.7e-03  &  0.99937  &  3.488e-04 \\ 
EUR-GBP  &  0.948  & $\pm$  2.6e-03  &  9.827e-01  & $\pm$ 6.2e-03  &  0.99818  &  5.941e-04 \\ 
EUR-JPY  &  0.961  & $\pm$  1.7e-03  &  9.921e-01  & $\pm$ 4.0e-03  &  0.99926  &  4.974e-04 \\ 
EUR-USD  &  0.959  & $\pm$  1.9e-03  &  9.918e-01  & $\pm$ 4.6e-03  &  0.99902  &  6.268e-04 \\ 
GBP-CHF  &  0.961  & $\pm$  1.6e-03  &  9.918e-01  & $\pm$ 3.8e-03  &  0.99934  &  4.941e-04 \\ 
GBP-JPY  &  0.959  & $\pm$  1.7e-03  &  9.926e-01  & $\pm$ 4.1e-03  &  0.99922  &  5.991e-04 \\ 
GBP-USD  &  0.961  & $\pm$  1.5e-03  &  9.918e-01  & $\pm$ 3.6e-03  &  0.99940  &  4.778e-04 \\ 
GRW  &  0.937  & $\pm$  2.6e-03  &  9.943e-01  & $\pm$ 6.5e-03  &  0.99805  &  1.656e-03 \\ 
USD-CHF  &  0.958  & $\pm$  1.9e-03  &  9.913e-01  & $\pm$ 4.6e-03  &  0.99901  &  6.267e-04 \\ 
USD-JPY  &  0.954  & $\pm$  2.3e-03  &  9.934e-01  & $\pm$ 5.6e-03  &  0.99858  &  8.996e-04 \\ 
\hline \\ [-2.1ex]
Currency average & 0.95 & (8.6e-03) & 9.89e-01 & (4.2e-03) & & \\
\end{tabular}
\end{table}

\begin{table}
\caption{Overshoot move, law (\ref{slcondx1}), $*=os$}
\label{tbl:12}
\centering
\begin{tabular}{lcccccc} \\ [-2.5ex]
\hline  \\  [-2.1ex]
Currency & $E_{x,os}$ & $\Delta E_{x,os}$ & $ C_{x,os}$  & $\Delta  C_{x,os}$ & Adj. $R^2$&   $R^2_{quad} - R^2_{lin}$ \\   [0.5ex]
\hline \\ [-2.1ex]
AUD-JPY  &  1.081  & $\pm$  2.5e-03  &  1.012e+00  & $\pm$ 5.5e-03  &  0.99865  &  -4.901e-06 \\ 
AUD-USD  &  1.060  & $\pm$  2.8e-03  &  9.788e-01  & $\pm$ 6.1e-03  &  0.99823  &  1.273e-04 \\ 
CHF-JPY  &  1.064  & $\pm$  4.9e-03  &  1.098e+00  & $\pm$ 1.2e-02  &  0.99478  &  3.123e-03 \\ 
EUR-AUD  &  1.066  & $\pm$  3.8e-03  &  1.121e+00  & $\pm$ 9.3e-03  &  0.99692  &  -9.706e-06 \\ 
EUR-CHF  &  1.071  & $\pm$  6.0e-03  &  1.118e+00  & $\pm$ 1.5e-02  &  0.99228  &  3.963e-03 \\ 
EUR-GBP  &  1.103  & $\pm$  7.4e-03  &  1.164e+00  & $\pm$ 1.8e-02  &  0.98905  &  7.059e-03 \\ 
EUR-JPY  &  1.052  & $\pm$  2.9e-03  &  1.004e+00  & $\pm$ 6.4e-03  &  0.99815  &  1.110e-03 \\ 
EUR-USD  &  0.996  & $\pm$  2.2e-03  &  9.879e-01  & $\pm$ 5.1e-03  &  0.99880  &  1.930e-05 \\ 
GBP-CHF  &  1.041  & $\pm$  4.0e-03  &  1.168e+00  & $\pm$ 1.1e-02  &  0.99629  &  2.002e-03 \\ 
GBP-JPY  &  1.042  & $\pm$  2.7e-03  &  1.027e+00  & $\pm$ 6.3e-03  &  0.99830  &  3.072e-04 \\ 
GBP-USD  &  1.003  & $\pm$  2.0e-03  &  9.871e-01  & $\pm$ 4.6e-03  &  0.99900  &  1.002e-04 \\ 
GRW  &  0.945  & $\pm$  4.7e-03  &  9.812e-01  & $\pm$ 1.1e-02  &  0.99390  &  2.167e-03 \\ 
USD-CHF  &  0.990  & $\pm$  2.8e-03  &  1.045e+00  & $\pm$ 7.0e-03  &  0.99798  &  9.718e-04 \\ 
USD-JPY  &  0.987  & $\pm$  2.6e-03  &  1.067e+00  & $\pm$ 6.6e-03  &  0.99825  &  1.625e-04 \\ 
\hline \\ [-2.1ex]
Currency average & 1.04 & (3.8e-02) & 1.06e+00 & (6.8e-02) & & \\
\end{tabular}
\end{table}

\begin{table}
\caption{Time of total move, law (\ref{slcondx2}), $*=tm$}
\label{tbl:13}
\centering
\begin{tabular}{lcccccc} \\ [-2.5ex]
\hline  \\  [-2.1ex]
Currency & $E_{t,tm}$ & $\Delta E_{t,tm}$ & $ C_{t,tm}$  & $\Delta  C_{t,tm}$ & Adj. $R^2$&   $R^2_{quad} - R^2_{lin}$ \\   [0.5ex]
\hline \\ [-2.1ex]
AUD-JPY  &  2.049  & $\pm$  4.6e-03  &  2.450e-03  & $\pm$ 3.5e-05  &  0.99876  &  3.127e-04 \\ 
AUD-USD  &  1.950  & $\pm$  4.5e-03  &  1.616e-03  & $\pm$ 2.5e-05  &  0.99870  &  5.183e-04 \\ 
CHF-JPY  &  2.066  & $\pm$  4.8e-03  &  2.063e-03  & $\pm$ 3.1e-05  &  0.99868  &  1.143e-04 \\ 
EUR-AUD  &  2.133  & $\pm$  5.8e-03  &  2.534e-03  & $\pm$ 4.4e-05  &  0.99816  &  6.282e-05 \\ 
EUR-CHF  &  2.103  & $\pm$  8.5e-03  &  9.655e-04  & $\pm$ 2.9e-05  &  0.99594  &  2.089e-03 \\ 
EUR-GBP  &  2.167  & $\pm$  8.1e-03  &  1.926e-03  & $\pm$ 4.8e-05  &  0.99653  &  1.311e-03 \\ 
EUR-JPY  &  2.004  & $\pm$  2.8e-03  &  1.638e-03  & $\pm$ 1.6e-05  &  0.99952  &  8.980e-05 \\ 
EUR-USD  &  1.886  & $\pm$  3.9e-03  &  1.053e-03  & $\pm$ 1.6e-05  &  0.99892  &  3.705e-04 \\ 
GBP-CHF  &  2.124  & $\pm$  3.3e-03  &  1.919e-03  & $\pm$ 2.0e-05  &  0.99941  &  8.651e-05 \\ 
GBP-JPY  &  2.018  & $\pm$  3.4e-03  &  1.809e-03  & $\pm$ 2.0e-05  &  0.99931  &  -2.651e-06 \\ 
GBP-USD  &  1.902  & $\pm$  3.2e-03  &  1.026e-03  & $\pm$ 1.3e-05  &  0.99928  &  1.987e-04 \\ 
GRW  &  1.791  & $\pm$  9.3e-03  &  6.967e-03  & $\pm$ 2.0e-04  &  0.99338  &  4.870e-03 \\ 
USD-CHF  &  1.898  & $\pm$  3.1e-03  &  1.229e-03  & $\pm$ 1.4e-05  &  0.99934  &  1.430e-04 \\ 
USD-JPY  &  1.925  & $\pm$  4.6e-03  &  1.260e-03  & $\pm$ 2.1e-05  &  0.99857  &  6.307e-04 \\ 
\hline \\ [-2.1ex]
Currency average & 2.02 & (9.8e-02) & 1.65e-03 & (5.3e-04) & & \\
\end{tabular}
\end{table}

\begin{table}
\caption{Time of directional change, law (\ref{slcondx2}), $*=dc$}
\label{tbl:14}
\centering
\begin{tabular}{lcccccc} \\ [-2.5ex]
\hline  \\  [-2.1ex]
Currency & $E_{t,dc}$ & $\Delta E_{t,dc}$ & $ C_{t,dc}$  & $\Delta  C_{t,dc}$ & Adj. $R^2$&   $R^2_{quad} - R^2_{lin}$ \\   [0.5ex]
\hline \\ [-2.1ex]
AUD-JPY  &  1.996  & $\pm$  4.5e-03  &  3.654e-03  & $\pm$ 5.0e-05  &  0.99875  &  2.448e-04 \\ 
AUD-USD  &  1.871  & $\pm$  4.1e-03  &  2.280e-03  & $\pm$ 3.2e-05  &  0.99882  &  4.129e-04 \\ 
CHF-JPY  &  2.015  & $\pm$  5.5e-03  &  3.096e-03  & $\pm$ 5.3e-05  &  0.99812  &  5.073e-04 \\ 
EUR-AUD  &  2.102  & $\pm$  4.7e-03  &  3.789e-03  & $\pm$ 5.1e-05  &  0.99875  &  1.251e-04 \\ 
EUR-CHF  &  2.066  & $\pm$  7.1e-03  &  1.463e-03  & $\pm$ 3.5e-05  &  0.99709  &  2.023e-03 \\ 
EUR-GBP  &  2.135  & $\pm$  6.3e-03  &  2.838e-03  & $\pm$ 5.3e-05  &  0.99786  &  4.578e-04 \\ 
EUR-JPY  &  1.941  & $\pm$  4.1e-03  &  2.415e-03  & $\pm$ 3.3e-05  &  0.99887  &  6.149e-04 \\ 
EUR-USD  &  1.846  & $\pm$  3.4e-03  &  1.636e-03  & $\pm$ 2.1e-05  &  0.99915  &  3.813e-05 \\ 
GBP-CHF  &  2.099  & $\pm$  3.9e-03  &  2.874e-03  & $\pm$ 3.4e-05  &  0.99914  &  1.035e-04 \\ 
GBP-JPY  &  1.958  & $\pm$  3.1e-03  &  2.626e-03  & $\pm$ 2.7e-05  &  0.99938  &  1.054e-04 \\ 
GBP-USD  &  1.866  & $\pm$  2.4e-03  &  1.596e-03  & $\pm$ 1.4e-05  &  0.99960  &  6.737e-05 \\ 
GRW  &  1.774  & $\pm$  9.8e-03  &  1.235e-02  & $\pm$ 3.4e-04  &  0.99240  &  6.439e-03 \\ 
USD-CHF  &  1.888  & $\pm$  2.8e-03  &  2.016e-03  & $\pm$ 2.0e-05  &  0.99945  &  5.768e-05 \\ 
USD-JPY  &  1.914  & $\pm$  4.1e-03  &  2.057e-03  & $\pm$ 2.9e-05  &  0.99885  &  5.935e-04 \\ 
\hline \\ [-2.1ex]
Currency average & 1.98 & (1.0e-01) & 2.49e-03 & (7.5e-04) & & \\
\end{tabular}
\end{table}

\begin{table}
\caption{Time of overshoot, law (\ref{slcondx2}), $*=os$}
\label{tbl:15}
\centering
\begin{tabular}{lcccccc} \\ [-2.5ex]
\hline  \\  [-2.1ex]
Currency & $E_{t,os}$ & $\Delta E_{t,os}$ & $ C_{t,os}$  & $\Delta  C_{t,os}$ & Adj. $R^2$&   $R^2_{quad} - R^2_{lin}$ \\   [0.5ex]
\hline \\ [-2.1ex]
AUD-JPY  &  2.079  & $\pm$  5.4e-03  &  3.241e-03  & $\pm$ 5.2e-05  &  0.99831  &  2.681e-04 \\ 
AUD-USD  &  1.997  & $\pm$  5.5e-03  &  2.273e-03  & $\pm$ 4.1e-05  &  0.99810  &  3.617e-04 \\ 
CHF-JPY  &  2.093  & $\pm$  5.8e-03  &  2.698e-03  & $\pm$ 4.7e-05  &  0.99812  &  4.505e-05 \\ 
EUR-AUD  &  2.150  & $\pm$  6.8e-03  &  3.274e-03  & $\pm$ 6.4e-05  &  0.99750  &  2.799e-05 \\ 
EUR-CHF  &  2.119  & $\pm$  1.0e-02  &  1.234e-03  & $\pm$ 4.2e-05  &  0.99441  &  2.222e-03 \\ 
EUR-GBP  &  2.190  & $\pm$  1.1e-02  &  2.519e-03  & $\pm$ 7.8e-05  &  0.99420  &  2.342e-03 \\ 
EUR-JPY  &  2.035  & $\pm$  3.1e-03  &  2.182e-03  & $\pm$ 2.1e-05  &  0.99944  &  4.337e-06 \\ 
EUR-USD  &  1.906  & $\pm$  5.2e-03  &  1.411e-03  & $\pm$ 2.7e-05  &  0.99816  &  6.454e-04 \\ 
GBP-CHF  &  2.136  & $\pm$  4.9e-03  &  2.469e-03  & $\pm$ 3.7e-05  &  0.99869  &  1.094e-04 \\ 
GBP-JPY  &  2.049  & $\pm$  4.4e-03  &  2.435e-03  & $\pm$ 3.4e-05  &  0.99884  &  2.919e-05 \\ 
GBP-USD  &  1.921  & $\pm$  4.4e-03  &  1.372e-03  & $\pm$ 2.2e-05  &  0.99871  &  2.923e-04 \\ 
GRW  &  1.796  & $\pm$  9.6e-03  &  8.900e-03  & $\pm$ 2.5e-04  &  0.99295  &  3.823e-03 \\ 
USD-CHF  &  1.903  & $\pm$  4.0e-03  &  1.597e-03  & $\pm$ 2.3e-05  &  0.99891  &  2.088e-04 \\ 
USD-JPY  &  1.930  & $\pm$  5.5e-03  &  1.624e-03  & $\pm$ 3.2e-05  &  0.99796  &  6.314e-04 \\ 
\hline \\ [-2.1ex]
Currency average & 2.04 & (1.0e-01) & 2.18e-03 & (6.9e-04) & & \\
\end{tabular}
\end{table}

\begin{table}
\caption{Total-move tick count, law (\ref{slcondx3}), $*=tm$}
\label{tbl:16}
\centering
\begin{tabular}{lcccccc} \\ [-2.5ex]
\hline  \\  [-2.1ex]
Currency & $E_{\textsf{\tiny N},tm}$ & $\Delta E_{\textsf{\tiny N},tm}$ & $ C_{\textsf{\tiny N},tm}$  & $\Delta  C_{\textsf{\tiny N},tm}$ & Adj. $R^2$&   $R^2_{quad} - R^2_{lin}$ \\   [0.5ex]
\hline \\ [-2.1ex]
AUD-JPY  &  2.048  & $\pm$  4.6e-03  &  1.861e-02  & $\pm$ 1.9e-04  &  0.99877  &  3.049e-04 \\ 
AUD-USD  &  1.949  & $\pm$  4.4e-03  &  2.008e-02  & $\pm$ 2.1e-04  &  0.99873  &  4.784e-04 \\ 
CHF-JPY  &  2.063  & $\pm$  4.9e-03  &  1.661e-02  & $\pm$ 1.8e-04  &  0.99862  &  1.473e-04 \\ 
EUR-AUD  &  2.131  & $\pm$  5.8e-03  &  2.098e-02  & $\pm$ 2.6e-04  &  0.99813  &  5.135e-05 \\ 
EUR-CHF  &  2.121  & $\pm$  6.3e-03  &  2.166e-02  & $\pm$ 2.9e-04  &  0.99778  &  1.058e-03 \\ 
EUR-GBP  &  2.177  & $\pm$  7.7e-03  &  2.225e-02  & $\pm$ 3.5e-04  &  0.99687  &  9.997e-04 \\ 
EUR-JPY  &  2.003  & $\pm$  2.8e-03  &  1.864e-02  & $\pm$ 1.2e-04  &  0.99950  &  1.014e-04 \\ 
EUR-USD  &  1.893  & $\pm$  4.1e-03  &  1.931e-02  & $\pm$ 1.9e-04  &  0.99884  &  5.552e-04 \\ 
GBP-CHF  &  2.121  & $\pm$  3.4e-03  &  2.024e-02  & $\pm$ 1.5e-04  &  0.99937  &  1.176e-04 \\ 
GBP-JPY  &  2.016  & $\pm$  3.4e-03  &  1.909e-02  & $\pm$ 1.5e-04  &  0.99927  &  -2.765e-06 \\ 
GBP-USD  &  1.902  & $\pm$  3.2e-03  &  1.883e-02  & $\pm$ 1.4e-04  &  0.99931  &  2.122e-04 \\ 
GRW  &  1.792  & $\pm$  9.3e-03  &  1.767e-02  & $\pm$ 4.3e-04  &  0.99338  &  4.899e-03 \\ 
USD-CHF  &  1.898  & $\pm$  3.0e-03  &  1.866e-02  & $\pm$ 1.4e-04  &  0.99937  &  1.604e-04 \\ 
USD-JPY  &  1.922  & $\pm$  4.6e-03  &  2.047e-02  & $\pm$ 2.2e-04  &  0.99858  &  5.640e-04 \\ 
\hline \\ [-2.1ex]
Currency average & 2.02 & (1.0e-02) & 1.97e-02 & (1.5e-03) & & \\
\end{tabular}
\end{table}

\begin{table}
\caption{Directional-change tick count, law (\ref{slcondx3}), $*=dc$}
\label{tbl:17}
\centering
\begin{tabular}{lcccccc} \\ [-2.5ex]
\hline  \\  [-2.1ex]
Currency & $E_{\textsf{\tiny N},dc}$ & $\Delta E_{\textsf{\tiny N},dc}$ & $ C_{\textsf{\tiny N},dc}$  & $\Delta  C_{\textsf{\tiny N},dc}$ & Adj. $R^2$&   $R^2_{quad} - R^2_{lin}$ \\   [0.5ex]
\hline \\ [-2.1ex]
AUD-JPY  &  2.002  & $\pm$  7.5e-03  &  2.974e-02  & $\pm$ 4.7e-04  &  0.99647  &  7.871e-04 \\ 
AUD-USD  &  1.904  & $\pm$  7.8e-03  &  3.375e-02  & $\pm$ 5.7e-04  &  0.99583  &  2.323e-04 \\ 
CHF-JPY  &  2.009  & $\pm$  8.1e-03  &  2.652e-02  & $\pm$ 4.6e-04  &  0.99600  &  -4.305e-06 \\ 
EUR-AUD  &  2.128  & $\pm$  4.6e-03  &  3.406e-02  & $\pm$ 3.0e-04  &  0.99884  &  9.624e-06 \\ 
EUR-CHF  &  2.198  & $\pm$  1.0e-02  &  3.896e-02  & $\pm$ 7.4e-04  &  0.99442  &  2.203e-03 \\ 
EUR-GBP  &  2.205  & $\pm$  1.5e-02  &  3.748e-02  & $\pm$ 1.0e-03  &  0.98832  &  1.535e-03 \\ 
EUR-JPY  &  2.040  & $\pm$  9.4e-03  &  3.465e-02  & $\pm$ 6.6e-04  &  0.99471  &  1.755e-03 \\ 
EUR-USD  &  2.017  & $\pm$  1.2e-02  &  4.187e-02  & $\pm$ 1.0e-03  &  0.99088  &  2.739e-03 \\ 
GBP-CHF  &  2.181  & $\pm$  6.4e-03  &  3.548e-02  & $\pm$ 4.2e-04  &  0.99786  &  7.460e-04 \\ 
GBP-JPY  &  2.048  & $\pm$  5.2e-03  &  3.436e-02  & $\pm$ 3.6e-04  &  0.99839  &  5.759e-04 \\ 
GBP-USD  &  1.985  & $\pm$  7.0e-03  &  3.782e-02  & $\pm$ 5.3e-04  &  0.99694  &  9.365e-04 \\ 
GRW  &  1.809  & $\pm$  6.6e-03  &  3.350e-02  & $\pm$ 5.0e-04  &  0.99669  &  2.568e-03 \\ 
USD-CHF  &  1.991  & $\pm$  9.4e-03  &  3.772e-02  & $\pm$ 7.2e-04  &  0.99451  &  7.606e-04 \\ 
USD-JPY  &  2.052  & $\pm$  9.9e-03  &  4.266e-02  & $\pm$ 8.1e-04  &  0.99427  &  9.073e-04 \\ 
\hline \\ [-2.1ex]
Currency average & 2.06 & (9.2e-02) & 3.58e-02 & (4.5e-03) & & \\
\end{tabular}
\end{table}

\begin{table}
\caption{Overshoot tick count, law (\ref{slcondx3}), $*=os$}
\label{tbl:18}
\centering
\begin{tabular}{lcccccc} \\ [-2.5ex]
\hline  \\  [-2.1ex]
Currency & $E_{\textsf{\tiny N},os}$ & $\Delta E_{\textsf{\tiny N},os}$ & $ C_{\textsf{\tiny N},os}$  & $\Delta  C_{\textsf{\tiny N},os}$ & Adj. $R^2$&   $R^2_{quad} - R^2_{lin}$ \\   [0.5ex]
\hline \\ [-2.1ex]
AUD-JPY  &  2.081  & $\pm$  4.7e-03  &  2.395e-02  & $\pm$ 2.4e-04  &  0.99873  &  5.969e-05 \\ 
AUD-USD  &  1.978  & $\pm$  4.5e-03  &  2.573e-02  & $\pm$ 2.5e-04  &  0.99873  &  4.303e-04 \\ 
CHF-JPY  &  2.101  & $\pm$  6.0e-03  &  2.140e-02  & $\pm$ 2.7e-04  &  0.99799  &  5.063e-04 \\ 
EUR-AUD  &  2.137  & $\pm$  7.9e-03  &  2.602e-02  & $\pm$ 4.2e-04  &  0.99658  &  3.846e-05 \\ 
EUR-CHF  &  2.089  & $\pm$  7.9e-03  &  2.554e-02  & $\pm$ 4.2e-04  &  0.99642  &  1.126e-03 \\ 
EUR-GBP  &  2.182  & $\pm$  8.9e-03  &  2.743e-02  & $\pm$ 4.8e-04  &  0.99587  &  1.462e-03 \\ 
EUR-JPY  &  1.998  & $\pm$  4.0e-03  &  2.246e-02  & $\pm$ 2.0e-04  &  0.99902  &  -8.981e-07 \\ 
EUR-USD  &  1.868  & $\pm$  6.6e-03  &  2.277e-02  & $\pm$ 3.6e-04  &  0.99687  &  1.731e-03 \\ 
GBP-CHF  &  2.095  & $\pm$  6.0e-03  &  2.416e-02  & $\pm$ 3.0e-04  &  0.99796  &  7.889e-05 \\ 
GBP-JPY  &  2.006  & $\pm$  5.3e-03  &  2.306e-02  & $\pm$ 2.7e-04  &  0.99828  &  4.535e-05 \\ 
GBP-USD  &  1.878  & $\pm$  5.2e-03  &  2.239e-02  & $\pm$ 2.8e-04  &  0.99806  &  7.896e-04 \\ 
GRW  &  1.783  & $\pm$  1.1e-02  &  2.194e-02  & $\pm$ 6.2e-04  &  0.99034  &  5.591e-03 \\ 
USD-CHF  &  1.871  & $\pm$  4.5e-03  &  2.220e-02  & $\pm$ 2.4e-04  &  0.99855  &  4.614e-04 \\ 
USD-JPY  &  1.884  & $\pm$  6.8e-03  &  2.405e-02  & $\pm$ 3.8e-04  &  0.99678  &  1.210e-03 \\ 
\hline \\ [-2.1ex]
Currency average & 2.01 & (1.1e-01) & 2.39e-02 & (1.8e-03) & & \\
\end{tabular}
\end{table}

\begin{table}
\caption{Cumulative total move (coastline), law (\ref{slcoast}), $*=tm$}
\label{tbl:19}
\centering
\begin{tabular}{lcccccc} \\ [-2.5ex]
\hline  \\  [-2.1ex]
Currency & $E_{cum,tm}$ & $\Delta E_{cum,tm}$ & $ C_{cum,tm}$  & $\Delta  C_{cum,tm}$ & Adj. $R^2$&   $R^2_{quad} - R^2_{lin}$ \\   [0.5ex]
\hline \\ [-2.1ex]
AUD-JPY  &  -1.048  & $\pm$  2.6e-03  &  2.139e+02  & $\pm$ 3.1e+00  &  0.99850  &  1.949e-04 \\ 
AUD-USD  &  -0.961  & $\pm$  2.7e-03  &  2.861e+02  & $\pm$ 4.9e+00  &  0.99805  &  1.054e-03 \\ 
CHF-JPY  &  -1.076  & $\pm$  2.5e-03  &  1.154e+02  & $\pm$ 1.4e+00  &  0.99868  &  1.307e-04 \\ 
EUR-AUD  &  -1.140  & $\pm$  3.7e-03  &  9.003e+01  & $\pm$ 1.5e+00  &  0.99738  &  -7.231e-06 \\ 
EUR-CHF  &  -1.189  & $\pm$  5.5e-03  &  1.312e+01  & $\pm$ 2.1e-01  &  0.99461  &  1.357e-04 \\ 
EUR-GBP  &  -1.184  & $\pm$  4.9e-03  &  3.711e+01  & $\pm$ 6.6e-01  &  0.99579  &  1.347e-03 \\ 
EUR-JPY  &  -1.005  & $\pm$  1.6e-03  &  1.601e+02  & $\pm$ 1.4e+00  &  0.99938  &  2.140e-04 \\ 
EUR-USD  &  -0.937  & $\pm$  3.8e-03  &  2.009e+02  & $\pm$ 4.8e+00  &  0.99583  &  2.997e-03 \\ 
GBP-CHF  &  -1.145  & $\pm$  2.1e-03  &  5.351e+01  & $\pm$ 4.5e-01  &  0.99916  &  -3.260e-06 \\ 
GBP-JPY  &  -1.024  & $\pm$  1.9e-03  &  1.605e+02  & $\pm$ 1.7e+00  &  0.99910  &  4.353e-07 \\ 
GBP-USD  &  -0.929  & $\pm$  2.4e-03  &  1.878e+02  & $\pm$ 2.8e+00  &  0.99832  &  9.695e-04 \\ 
GRW  &  -0.868  & $\pm$  6.8e-03  &  6.157e+02  & $\pm$ 3.3e+01  &  0.98509  &  1.284e-02 \\ 
USD-CHF  &  -0.939  & $\pm$  2.7e-03  &  2.515e+02  & $\pm$ 4.3e+00  &  0.99796  &  1.580e-03 \\ 
USD-JPY  &  -0.963  & $\pm$  3.3e-03  &  1.921e+02  & $\pm$ 3.8e+00  &  0.99701  &  2.282e-03 \\ 
hline \\ [-2.1ex]
Currency average & -1.04 & (9.7e-02) & 1.51e+02 & (8.4e+01) & & \\
\end{tabular}
\end{table}

\begin{table}
\caption{Cumulative  cost-adjusted total move, law (\ref{slcoast}), $*=tm$, fitted from $0.2\%$}
\label{tbl:20}
\centering
\begin{tabular}{lcccccc} \\ [-2.5ex]
\hline  \\  [-2.1ex]
Currency & $E_{cum,tm}$ & $\Delta E_{cum,tm}$ & $ C_{cum,tm}$  & $\Delta  C_{cum,tm}$ & Adj. $R^2$&   $R^2_{quad} - R^2_{lin}$ \\   [0.5ex]
\hline \\ [-2.1ex]
AUD-JPY  &  -0.941  & $\pm$  2.9e-03  &  3.378e+02  & $\pm$ 6.2e+00  &  0.99874  &  3.358e-05 \\ 
AUD-USD  &  -0.943  & $\pm$  2.0e-03  &  2.920e+02  & $\pm$ 3.6e+00  &  0.99940  &  7.101e-05 \\ 
CHF-JPY  &  -0.956  & $\pm$  3.4e-03  &  1.854e+02  & $\pm$ 3.5e+00  &  0.99838  &  1.087e-03 \\ 
EUR-AUD  &  -0.970  & $\pm$  4.9e-03  &  1.652e+02  & $\pm$ 4.3e+00  &  0.99678  &  1.822e-03 \\ 
EUR-CHF  &  -1.201  & $\pm$  2.0e-02  &  1.254e+01  & $\pm$ 5.5e-01  &  0.96693  &  1.789e-02 \\ 
EUR-GBP  &  -1.034  & $\pm$  7.9e-03  &  5.743e+01  & $\pm$ 1.8e+00  &  0.99249  &  3.734e-03 \\ 
EUR-JPY  &  -0.915  & $\pm$  2.0e-03  &  2.454e+02  & $\pm$ 2.9e+00  &  0.99941  &  1.810e-04 \\ 
EUR-USD  &  -0.980  & $\pm$  8.2e-03  &  1.524e+02  & $\pm$ 6.5e+00  &  0.99102  &  5.138e-03 \\ 
GBP-CHF  &  -1.059  & $\pm$  7.4e-03  &  6.783e+01  & $\pm$ 2.1e+00  &  0.99366  &  3.061e-03 \\ 
GBP-JPY  &  -0.912  & $\pm$  2.3e-03  &  2.666e+02  & $\pm$ 3.8e+00  &  0.99919  &  9.254e-06 \\ 
GBP-USD  &  -0.916  & $\pm$  4.2e-03  &  1.903e+02  & $\pm$ 4.6e+00  &  0.99730  &  7.322e-04 \\ 
GRW  &  -0.994  & $\pm$  9.2e-03  &  2.588e+02  & $\pm$ 1.4e+01  &  0.98897  &  3.848e-03 \\ 
USD-CHF  &  -0.949  & $\pm$  5.4e-03  &  2.214e+02  & $\pm$ 6.9e+00  &  0.99582  &  3.302e-03 \\ 
USD-JPY  &  -0.979  & $\pm$  4.4e-03  &  1.650e+02  & $\pm$ 3.8e+00  &  0.99744  &  1.522e-03 \\ 
\hline \\ [-2.1ex]
Currency average & -0.98 & (7.9e-02) & 1.82e+02 & (9.5e+01) & & \\  [1.0ex]
\end{tabular}
(For the GRW a constant spread of $0.02\%$ was introduced.)
\end{table}

\begin{table}
\caption{Cumulative directional change, law (\ref{slcoast}), $*=dc$}
\label{tbl:21}
\centering
\begin{tabular}{lcccccc} \\ [-2.5ex]
\hline  \\  [-2.1ex]
Currency & $E_{cum,dc}$ & $\Delta E_{cum,dc}$ & $ C_{cum,dc}$  & $\Delta  C_{cum,dc}$ & Adj. $R^2$&   $R^2_{quad} - R^2_{lin}$ \\   [0.5ex]
\hline \\ [-2.1ex]
AUD-JPY  &  -1.108  & $\pm$  2.8e-03  &  8.654e+01  & $\pm$ 1.1e+00  &  0.99843  &  2.727e-05 \\ 
AUD-USD  &  -1.014  & $\pm$  2.3e-03  &  1.075e+02  & $\pm$ 1.3e+00  &  0.99867  &  2.578e-04 \\ 
CHF-JPY  &  -1.121  & $\pm$  4.0e-03  &  5.387e+01  & $\pm$ 9.0e-01  &  0.99678  &  1.674e-03 \\ 
EUR-AUD  &  -1.191  & $\pm$  4.8e-03  &  4.376e+01  & $\pm$ 7.9e-01  &  0.99590  &  1.308e-05 \\ 
EUR-CHF  &  -1.232  & $\pm$  6.2e-03  &  7.196e+00  & $\pm$ 1.1e-01  &  0.99367  &  1.967e-04 \\ 
EUR-GBP  &  -1.240  & $\pm$  7.0e-03  &  1.940e+01  & $\pm$ 4.1e-01  &  0.99213  &  4.493e-03 \\ 
EUR-JPY  &  -1.044  & $\pm$  3.0e-03  &  6.849e+01  & $\pm$ 9.5e-01  &  0.99797  &  1.399e-03 \\ 
EUR-USD  &  -0.954  & $\pm$  3.8e-03  &  8.805e+01  & $\pm$ 1.8e+00  &  0.99597  &  1.953e-03 \\ 
GBP-CHF  &  -1.178  & $\pm$  3.1e-03  &  2.851e+01  & $\pm$ 3.0e-01  &  0.99831  &  5.737e-04 \\ 
GBP-JPY  &  -1.061  & $\pm$  2.9e-03  &  7.116e+01  & $\pm$ 9.5e-01  &  0.99813  &  3.460e-04 \\ 
GBP-USD  &  -0.948  & $\pm$  2.2e-03  &  8.117e+01  & $\pm$ 9.2e-01  &  0.99872  &  2.150e-04 \\ 
GRW  &  -0.874  & $\pm$  7.6e-03  &  2.630e+02  & $\pm$ 1.4e+01  &  0.98147  &  1.385e-02 \\ 
USD-CHF  &  -0.954  & $\pm$  1.9e-03  &  1.143e+02  & $\pm$ 1.2e+00  &  0.99901  &  1.745e-04 \\ 
USD-JPY  &  -0.978  & $\pm$  2.7e-03  &  9.014e+01  & $\pm$ 1.3e+00  &  0.99811  &  7.933e-04 \\ 
\hline \\ [-2.1ex]
Currency average & -1.08 & (1.1e-01) & 6.62e+01 & (3.4e+01) & & \\
\end{tabular}
\end{table}

\begin{table}
\caption{Cumulative overshoot, law (\ref{slcoast}), $*=os$}
\label{tbl:22}
\centering
\begin{tabular}{lcccccc} \\ [-2.5ex]
\hline  \\  [-2.1ex]
Currency & $E_{cum,os}$ & $\Delta E_{cum,os}$ & $ C_{cum,os}$  & $\Delta  C_{cum,os}$ & Adj. $R^2$&   $R^2_{quad} - R^2_{lin}$ \\   [0.5ex]
\hline \\ [-2.1ex]
AUD-JPY  &  -0.968  & $\pm$  3.5e-03  &  1.609e+02  & $\pm$ 3.2e+00  &  0.99678  &  1.365e-03 \\ 
AUD-USD  &  -0.892  & $\pm$  4.3e-03  &  2.054e+02  & $\pm$ 5.7e+00  &  0.99436  &  4.187e-03 \\ 
CHF-JPY  &  -1.006  & $\pm$  3.3e-03  &  7.612e+01  & $\pm$ 1.2e+00  &  0.99731  &  1.683e-03 \\ 
EUR-AUD  &  -1.071  & $\pm$  3.2e-03  &  5.888e+01  & $\pm$ 8.3e-01  &  0.99775  &  3.374e-04 \\ 
EUR-CHF  &  -1.124  & $\pm$  6.7e-03  &  7.751e+00  & $\pm$ 1.4e-01  &  0.99117  &  3.547e-03 \\ 
EUR-GBP  &  -1.084  & $\pm$  3.6e-03  &  2.503e+01  & $\pm$ 3.3e-01  &  0.99729  &  9.155e-04 \\ 
EUR-JPY  &  -0.954  & $\pm$  1.7e-03  &  1.010e+02  & $\pm$ 9.3e-01  &  0.99921  &  3.324e-04 \\ 
EUR-USD  &  -0.917  & $\pm$  4.2e-03  &  1.061e+02  & $\pm$ 2.6e+00  &  0.99474  &  4.471e-03 \\ 
GBP-CHF  &  -1.098  & $\pm$  3.4e-03  &  3.117e+01  & $\pm$ 4.0e-01  &  0.99766  &  1.305e-03 \\ 
GBP-JPY  &  -0.978  & $\pm$  2.4e-03  &  9.847e+01  & $\pm$ 1.2e+00  &  0.99852  &  5.018e-04 \\ 
GBP-USD  &  -0.906  & $\pm$  3.3e-03  &  9.983e+01  & $\pm$ 1.9e+00  &  0.99667  &  2.486e-03 \\ 
GRW  &  -0.866  & $\pm$  6.6e-03  &  2.803e+02  & $\pm$ 1.3e+01  &  0.98589  &  1.256e-02 \\ 
USD-CHF  &  -0.922  & $\pm$  4.5e-03  &  1.277e+02  & $\pm$ 3.3e+00  &  0.99413  &  5.365e-03 \\ 
USD-JPY  &  -0.945  & $\pm$  4.7e-03  &  9.785e+01  & $\pm$ 2.5e+00  &  0.99380  &  5.314e-03 \\ 
\hline \\ [-2.1ex]
Currency average & -0.99 & (8.0e-02) & 9.20e+01 & (5.5e+01) & & \\
\end{tabular}
\end{table}

\end{appendix}

\end{document}